\documentclass[aoas,preprint]{imsart}

\RequirePackage{amsthm, amsmath, amsfonts, amssymb}
\RequirePackage[authoryear]{natbib}
\RequirePackage[colorlinks, citecolor=blue, urlcolor=blue]{hyperref}
\RequirePackage{graphicx}
\RequirePackage{ulem,soul}
\usepackage{comment,xcolor}
\usepackage{subcaption}
\startlocaldefs

\theoremstyle{remark}
\newtheorem*{remark}{Remark}

\let\hat\widehat
\let\tilde\widetilde

\newcommand{\E}{\mbox{$\mathbb{E}$}}
\def\red{\textcolor{black}}

\usepackage{tikz}

\usetikzlibrary{positioning,shapes.geometric,calc,shapes,arrows.meta,arrows,decorations.markings, external, trees}
\tikzstyle{Arrow} = [
         thick,
         decoration={
                 markings,
                 mark=at position 1 with {
                        \arrow[thick]{latex}
} },
         shorten >= 3pt, preaction = {decorate}
        ]

\endlocaldefs

\begin{document}

\begin{frontmatter}
\title{Causal Inference in the Time of Covid-19 \thanksref{T1}}
\runtitle{Causal Inference in the Time of Covid}
\thankstext{T1}{Ventura and Wasserman are members of the Delphi Group at CMU \url{delphi.cmu.edu}.
This project arose from their work with Delphi. We are grateful for their help and support.
We thank Rob Tibshirani and the reviewers for suggestions that greatly improved the paper.
All the data can be obtained from the Delphi website \url{covidcast.cmu.edu}.}

\begin{aug}
\author[A]{\fnms{Matteo} \snm{Bonvini} \ead[label=e1, mark]{mbonvini@stat.cmu.edu}},
\author[A]{\fnms{Edward H.} \snm{Kennedy} \ead[label=e2,mark]{edward@stat.cmu.edu}}
\author[A, B]{\fnms{Valerie} \snm{Ventura} \ead[label=e3,mark]{vventura@stat.cmu.edu}}
\and
\author[A, B]{\fnms{Larry} \snm{Wasserman} \ead[label=e4,mark]{larry@stat.cmu.edu}} 
\address[A]{Department of Statistics and Data Science, Carnegie Mellon University}
\address[B]{Delphi Group, Carnegie Mellon University } 
{\printead{e1,e2,e3,e4}}
\end{aug}

\begin{center}
August 10 2021
\end{center}

\begin{abstract}
In this paper we develop
statistical methods
for causal inference
in epidemics.
Our focus is in estimating
the effect of social mobility on
deaths in the first year of the Covid-19 pandemic.
We propose a 
marginal structural model
motivated
by a bbasic epidemic model.
We estimate the counterfactual
time series of deaths under interventions
on mobility.
We conduct several types of sensitivity analyses.
We find that the data support the idea that
reduced mobility causes reduced deaths, 
but the conclusion comes with caveats.
There is evidence of sensitivity
to model misspecification and
unmeasured confounding
which implies that the size of the causal
effect needs to be interpreted with caution.
While there is little doubt the effect is real,
our work highlights the challenges
in drawing causal inferences from pandemic data.
\end{abstract}


\end{frontmatter}

\section{Introduction}

During a pandemic,
it is reasonable to expect
that reduced social mobility
will lead to fewer deaths.
But how do we quantify this effect?
In this paper
we combine 
\red{ideas from}
mechanistic
epidemic models with
modern causal inference tools
to answer this question
using state level data on deaths 
and mobility. 
Our goal is not to provide definitive estimates
for the effects but rather to
develop some methods and
highlight the challenges
in doing causal inference for pandemics.
We also show how a generative epidemic model
motivates a semiparametric causal model.

\red{
We use state death data at the weekly level.
The data are available at the daily county level
but the weekly state level data are more reliable. 
Indeed, the data are subject to many reporting issues.
It is not uncommon for a state to fail to report
many deaths for a few days and then suddenly
report a bunch of unreported deaths on a single day.
The problems are worse at the county level.
Also, there are many small counties with very little data.
We find using weekly state level data to be a good
compromise between the quantity and quality of the data.
We also note that epidemic analyses, such as flu
surveillance, are generally done
at the weekly level.}

Epidemics are usually modeled by using 
\red{generative models}, which fully specify the distribution of the
outcome (deaths). 
The most
common epidemic models relate exposure, infections, recoveries and
deaths by way of a set of differential equations.  The simplest
version is the SIR model (susceptible, infected, recovered) but there
are many flavors of the model.  We review the basic model in Section
\ref{section::models}.

Instead of a generative model,
we use a
marginal structural model (MSM)
(\cite{robins2000marginal,robins2000marginal2}).
An MSM is a semiparametric model that
directly models the effect of mobility on death
without specifying a generative model. 
Because it is semiparametric, it makes
fewer assumptions than a generative model.
However, our MSM is motivated by a modified SIR-type generative model. 

We model deaths in each state separately to reduce 
confounding due to state differences.
After obtaining model parameter estimates for each state,
\red{
we will be interested in the causal question:
what would happen if we set mobility to a certain value?
For example, how many deaths would have occurred
if mobility
had been reduced earlier, or if 
people had remained more vigilant throughout?
We follow standard causal language and refer to 
changing mobility as an intervention.
A different notion of intervention would be
a policy change like closing schools.
In this case, mobility is a mediator
meaning that the intervention affects the outcome through mobility.
In this paper we focus
on the effect of mobility on deaths
and refer to hypothetically setting mobility to a certain value
as an intervention.
Providing estimates of the effect
of mobility on deaths
is valuable 
so that we can tell policy makers 
what mobility level they should aim for
with their interventions.
Analyzing the effect of interventions
is also of interest
but in this paper we focus on the effect of mobility on deaths.}

We will see that the data provide
evidence for an effect of mobility.
But the data are very limited.
As mentioned above, we use state-specific models
with weekly resolution due to concerns about data quality
and unmeasured confounding due to geographic differences.
The result is that we have
about 40 observations per state.
With so little data,
we are restricted to use fairly simple models.
We do find significant causal effects
but we conduct sensitivity analyses
that show that the effects
need to be interpreted cautiously.
This sensitivity analysis includes
assessing the impact of
model assumptions and unobserved confounding.

\bigskip

{\bf Related Work.}
A number of researchers have
considered modeling the effect of 
causal interventions 
\red{(such as mobility and masks)} on
Covid-19.
Notable examples are
\cite{unwin2020state}, 
\cite{chang2020mobility},
and
\cite{ihme2020modeling}.
These authors develop very
detailed epidemic models
of the dynamics
of the disease.
One advantage of such an approach
is that one can then
consider the effects of a large array of 
potential interventions.
Further, the models themselves are of
great interest for understanding the 
\red{dynamics} of Covid-19.
However, these models are very complex, and they
involve a large number of parameters
including parameters for various latent variables.
Fitting such models
and assessing uncertainty is challenging.
Some authors take a Bayesian approach
with informative priors.
Others use heuristics
such as reporting intervals based on
using various settings of the parameters.
To the best of our knowledge,
it is not known how to get
valid, frequentist confidence intervals
in these complex models.
This is not meant as a criticism
of these papers
but rather, this reflects the intrinsic difficulty
of dealing with such models.
\red{Furthermore, 
when used for causal analysis,
parametrically specified epidemic models
are susceptible
to a problem known as the null paradox
which we discuss in Section 4.2.}

In contrast,
our goal is to make the model
as simple as possible and
to use standard estimating equation methods
so that standard errors can be
obtained fairly easily.
We do not claim that our approach is superior
but we do believe that
the model and the resulting
confidence intervals are
more transparent.
Getting precise results from our simple model
turns out to be challenging and
raises doubts about the accuracy
of published studies using highly
complex models.

The papers by
\cite{chernozhukov2020causal} and
\cite{xiong2020mobile}
are much closer to ours.
The authors of
\cite{chernozhukov2020causal} 
use a set of causal linear structural
equations
to model
weekly cases
as a function of social behavior (mobility)
and social behavior as a function of policies.
They model several policies
simultaneously and
they model all states simultaneously.
They do obtain valid frequentist confidence intervals.
\cite{xiong2020mobile}
construct a measure of mobility inflow
and using daily county level cases
they fit
a linear structural model
to relate cases
to mobility inflow.
Our approach differs in several ways:
we model deaths,
we focus only on the effect of mobility,
we model one state at a time,
and we use a MSM
rather than a generative model.
By modeling within each state, we have much less data at our disposal,
which makes modeling challenging.
On the other hand, 
the threat of confounding
due to state differences
is reduced.
By using a marginal structural model,
our approach is semiparametric
and so makes fewer assumptions.
Unlike these authors,
we focus on deaths instead of cases
because we find the data on cases to be quite unreliable in general; for example,
the availability of testing changed over time in various ways within and across states. 
\red{Moreover, the data early in the pandemic are very important and 
this is when case data were least reliable.}
Also, we place a strong emphasis on sensitivity analysis.
These analyses
complement each other nicely.

\bigskip

{\bf Paper Outline.}
We describe the data in
Section \ref{section::data}.
In Section \ref{section::causal}
we review some basics of causal inference.
In Section
\ref{section::models}
we construct the models
that we will use
and we explain how the models are fit
in Section \ref{section::fitting}.
The results are presented in Section
\ref{section::results}.
Concluding remarks
are in Section \ref{section::conclusion}.

\section{Data}
\label{section::data}

As mentioned earlier, we model each state separately, at the weekly level.
The data for each state have the form
$$
(A_1,Y_1),\ldots, (A_T,Y_T)
$$
where $A_t$ is mobility on week $t$
and $Y_t$ is the number of deaths due to Covid-19
on week $t$.
We obtained our data from
CMU's Delphi group (\url{cmu.covidcast.edu})
which gets the death data from
Johns Hopkins 
(\url{https://coronavirus.jhu.edu}) and
the mobility data from
Safegraph (\url{safegraph.com}).
The data are from 
Feb 15 2020 (week 1) to December 25 2020 (week 45).

\begin{figure}
	\begin{subfigure}[t]{.45\textwidth}
		\centering
		\includegraphics[width=\linewidth]{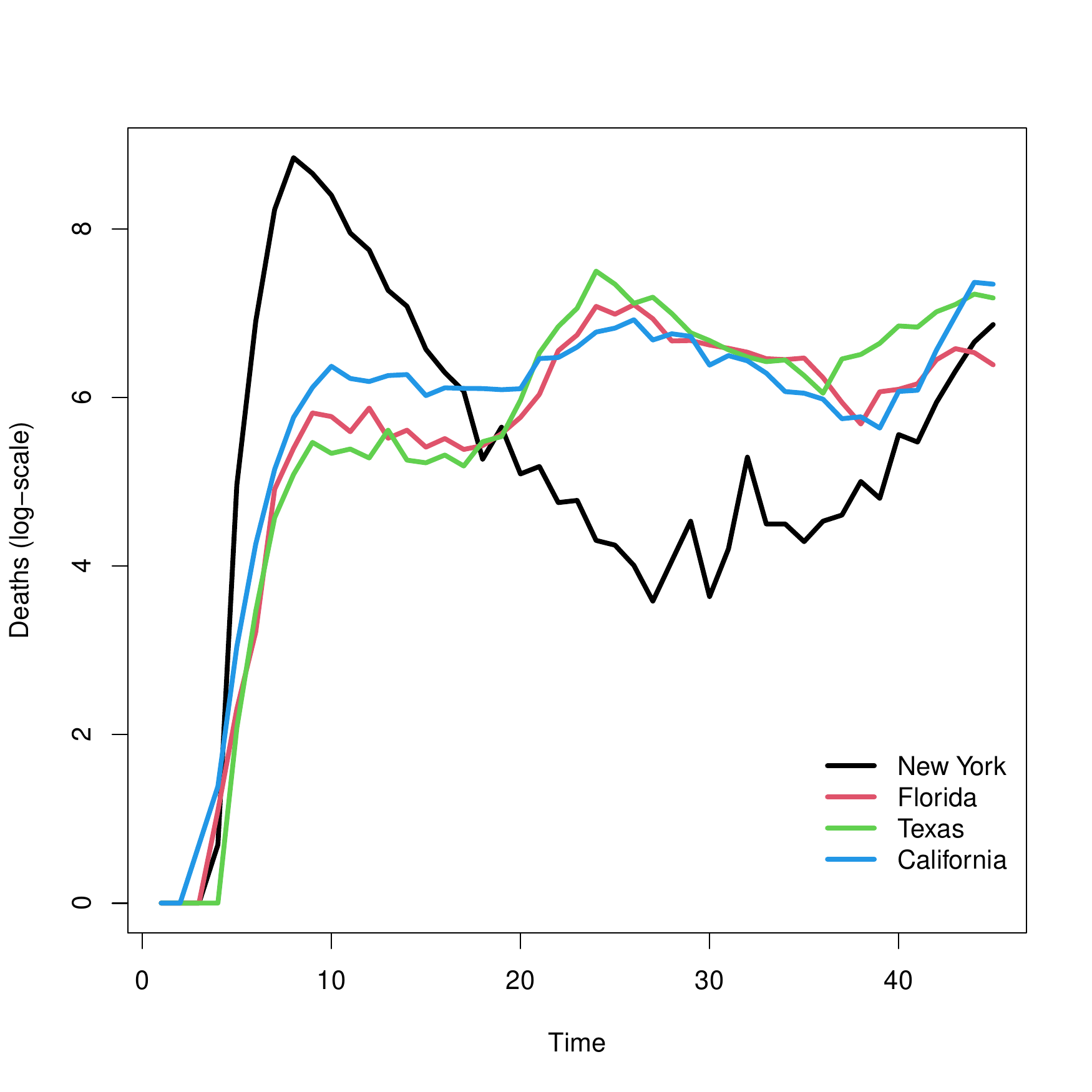}
		\caption{Plot of log deaths versus time (weeks), from
			Feb 15 2020 (week 1) to December 25 2020 (week 45), for four populous states.}
	\end{subfigure}\hfil
	\begin{subfigure}[t]{.45\textwidth}
		\centering
		\includegraphics[width=\linewidth]{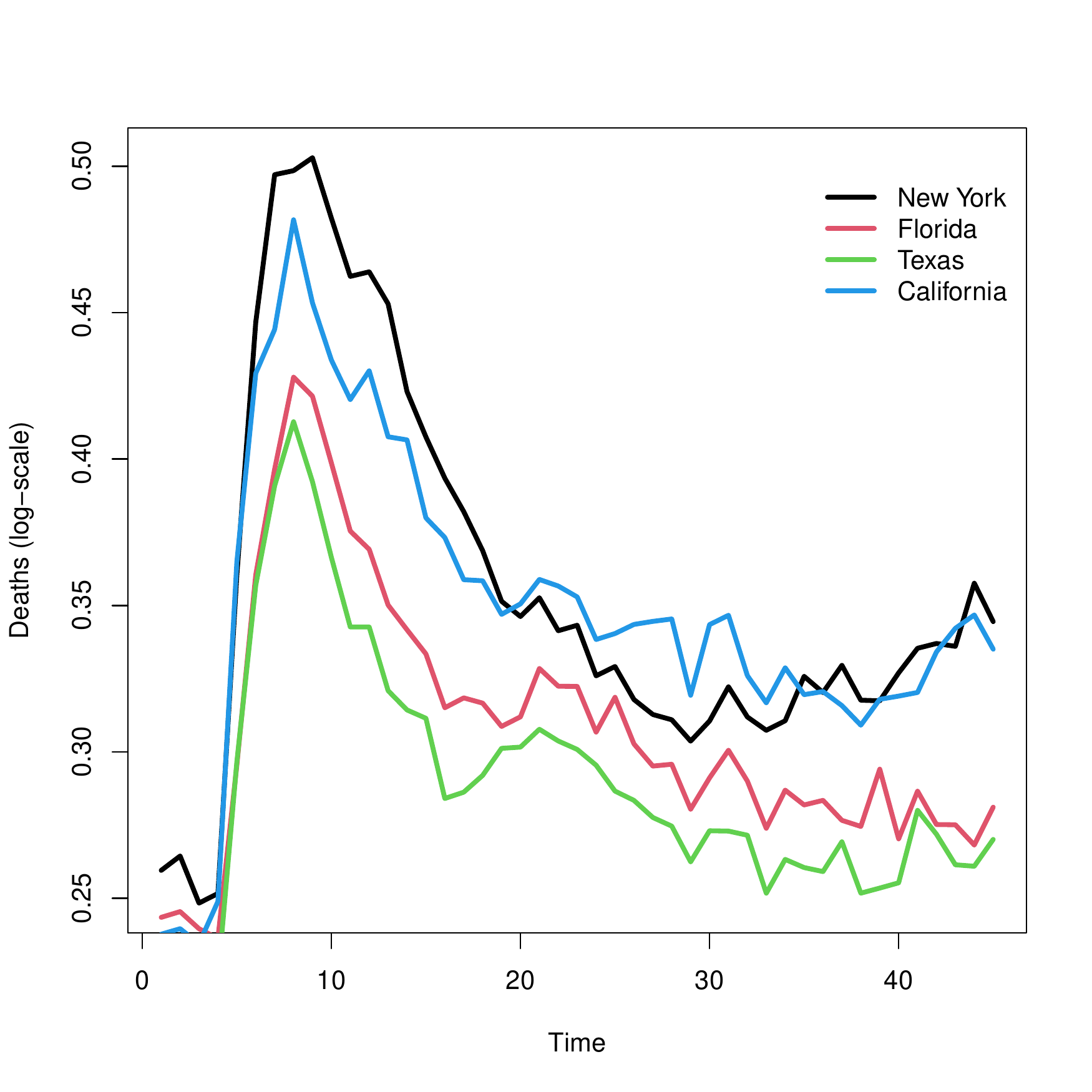}
		\caption{Plot of anti-mobility measure ``stay at home'' versus week.}
	\end{subfigure}
\caption{Plots of log deaths and anti-mobility across time. \label{fig::data}}
\end{figure}

Figure \ref{fig::data}
shows log deaths $L_t=\log(Y_t+1)$ and
``proportion at home'' $A_t$ which is
one of the mobility measures,
for four states. 
This is the fraction of mobile devices
that did not leave the immediate
area of their home.
In this case, a higher value means less mobility
so we can think of this measure as anti-mobility.
This is the variable we will use throughout.
In the rest of the paper we standardize mobility
by subtracting $A_1$ from each value of $A_t$
so that mobility starts at zero.

\section{Causal Inference}
\label{section::causal}

In this section, we
briefly
review basic ideas
from causal inference.
Consider weekly mobility and death data
$(A_1,Y_1),\ldots, (A_T,Y_T)$
in one state.
Define
$\overline{A}_t = (A_1,\ldots, A_t)$ and
$\overline{Y}_t = (Y_1,\ldots, Y_t)$
for $t\geq 1$.

Now consider
the causal question:
what would $Y_t$ be if we set
$\overline{A}_t$ equal to some value
$\overline{a}_t  =(a_1,\ldots, a_t)$?
Let
$Y^{\overline{a}_t}_t$
denote this counterfactual quantity.
It is important to distinguish
the observed data
$(\overline{A}_T,\overline{Y}_T)$
from the collection of
unobserved counterfactual random variables
$$
\Bigl\{ Y^{\overline{a}_T}:\ \overline{a}_T\in\mathbb{R}^T \Bigr\},
$$
which is an infinite collection of random vectors,
one for each possible mobility trajectory
$\overline{a}_T$.
We make the usual consistency assumption that
$\overline{Y}_T = \overline{Y}_T^{\overline{A}_T}$.
To make sure this is clear,
consider a simple case where 
a subject gets either treatment $A=1$ or control $A=0$.
In this case, 
the random variables are
$(A,Y,Y^0,Y^1)$
and the consistency assumption is that the observed outcome $Y$
satisfies $Y=Y^1$ if $A=1$ and
$Y=Y^0$ if $A=0$.

Causal inference
when the treatment varies over time
is subtle.
It may be tempting to simply regress $Y_T$ on the past 
and get the regression coefficient for mobility.
This strategy has serious problems
because
$\overline{Y}_{T-1}$
are both confounding  and mediating variables.
Indeed, previous deaths can affect both future mobility 
and future deaths, while also being affected by previous mobility.
More precisely, a large number of deaths implies
a large number of infections which can cause
future infections which then cause future deaths,
and a large number of deaths might
scare people into staying home.
So we must adjust for past deaths.
A common principle in epidemiology is to
adjust for pre-treatment variables
but not for post-treatment variables.
But $Y_s$ comes after $A_{s-1}$ and before $A_{s+1}$
making it both
a pre-treatment and post-treatment variable.
So how do we properly
define the causal effect?

The solution is to use Robins' $g$-formula.
Assuming for the moment that
there are no other confounding variables except past deaths,
\citet{robins1986new} proved that
the mean of 
$Y^{\overline{a}_t}_t$
is given by the g-formula:
\begin{equation}\label{eq::the-g}
\psi(\overline{a}_t)\equiv
\mathbb{E}[Y^{\overline{a}_t}_t] =
\int\cdots\int 
\mathbb{E}[Y_t | \overline{A}_t = \overline{a}_t, \overline{Y}_{t-1}=\overline{y}_{t-1}]
\prod_{s=1}^{t-1} p(y_s|\overline{y}_{s-1},\overline{a}_{s}) \ 
d y_{s};
\end{equation}
\red{$\psi(\overline{a}_t)$ is the causal effect we
seek to estimate.}
(We note that some authors denote
$\mathbb{E}[Y^{\overline{a}_t}_t]$ by
$\mathbb{E}[Y_t| {\rm do}(\overline{a}_t)]$.)
When there are other confounders $X_t$ besides past deaths, the
formula becomes
\begin{align*}
\psi(\overline{a}_t)  \equiv
\int\cdots\int 
\mathbb{E}[Y_t | \overline{A}_t = \overline{a}_t, \overline{Y}_{t-1}=\overline{y}_{t-1},\overline{X}_{t-1} = \overline{x}_{t-1}]
\prod_{s=1}^{t-1} p(y_s,x_s|\overline{y}_{s-1},\overline{a}_{s},\overline{x}_{s-1}) \ 
d y_{s}\, d x_s.
\end{align*}
Intuitively,
the $g$-formula can be obtained as follows.
The density
of $(\overline{y}_t,\overline{a}_t)$
can be written as
\begin{equation}\label{eq::joint}
p(\overline{y}_t,\overline{a}_t) =
\prod_{s=1}^t p(y_s|\overline{y}_{s-1},\overline{a}_{s}) p(a_s|\overline{a}_{s-1},\overline{y}_{s-1}).
\end{equation}
Now replace
$p(a_s|\overline{a}_{s-1},\overline{y}_{s-1})$
with a point mass at $a_s$
(i.e. the $A$'s are fixed, no longer random)
and then
find of the mean of $Y_t$
from this new distribution.
It will be useful later in the paper
to bear in mind that
$\psi(\overline{a}_t)\equiv \psi(\overline{a}_t,p)$
is a functional of the joint density $p$ from (\ref{eq::joint}).

\red{
For the causal effect $\psi(\overline{a}_t)$ to be identified
we require three standard assumptions.
These are:
(1) there is no unmeasured confounding.
Formally, this means that at each time,
the treatment is independent of the counterfactuals
given the past measured variables.
(2) The distribution of treatment has a positive density.
(3) Counterfactual consistency: If $\overline{A}_t=\overline{a_t}$ then
$Y_t = Y^{\overline{a}_t}$.
Later we add a fourth assumption, namely, that the dependence of mobility on the past satisfies
a Markov condition.
}

The next question is:
how do we estimate
$\psi(\overline{a}_t)$?
A natural idea
is to 
plug-in estimates
of all the unknown quantities in the $g$-formula
which leads to
\begin{equation}
\hat\psi(\overline{a}_t)\equiv
\int\cdots\int 
\hat{\mathbb{E}}[Y_t | \overline{A}_t = \overline{a}_t, \overline{Y}_{t-1}=\overline{y}_{t-1}]
\prod_{s=1}^{t-1} \hat{p}(y_s|\overline{y}_{s-1},\overline{a}_{s}) \ 
d y_{s}.
\end{equation}
As discussed in 
\cite{robins2000marginal,robins2000marginal2,robins1989analysis}
there are a number of problems
with this approach, called g-computation.
If we plug-in nonparametric estimates,
we quickly face the curse of dimensionality.
If we use parametric estimates,
we encounter the null-paradox
(\cite{rw}):
there may be no setting of the parameters which can represent the case where
there is no treatment effect, 
i.e.,  there is 
no setting of the parameters which makes $\psi(\overline{a}_t)$ a constant function of $\overline{a}_t$.
\red{We discuss the null paradox further in Section 4.2.}

An alternative approach
to estimating $\psi(\overline{a}_t)$
(\cite{robins2000marginal})
is to directly specify a 
parametric functional form 
$g(\overline{a}_t,\beta)$
for $\psi(\overline{a}_t)$.
Such a model is called a
marginal structural model (MSM).
\cite{robins2000marginal} showed that
$\beta$ can be estimated by solving the following {inverse-probability-weighted} estimating equation:
\begin{equation}\label{eq::ee}
\sum_t W_t^* \ h^*(\overline{A}_t)(Y_t - g(\overline{A}_t,\hat\beta)) =0,
\end{equation}
where the weights $W_t^*$ are defined by
\begin{equation}\label{eq::Wt0}
W_t^*= \prod_{s=1}^t  \frac{1}{\pi(A_s| \overline{A}_{s-1},\overline{Y}_{s-1})}
\end{equation} 
and $\pi(a_t|\cdot)$ is the conditional density of mobility, assumed to be positive.
\red{
We follow the common practice
(\cite{robins2000marginal})
of using stabilized weights
\begin{equation}\label{eq::Wt}
W_t= \prod_{s=1}^t  
\frac{\pi(A_s| \overline{A}_{s-1})}{\pi(A_s| \overline{A}_{s-1},\overline{Y}_{s-1})}.
\end{equation}
which corresponds to setting
$h^*(\overline{a}_t) = h(\overline{a}_t) \prod_{s=1}^t \pi(A_s \mid \overline{A}_{s-1})$
for some $h$. 
We discuss the choice of $h$
in Section \ref{section::fitting}.
We will then find it convenient to
rewrite (\ref{eq::ee}) as
\begin{equation}\label{eq::ee2}
\sum_t W_t \ h(\overline{A}_t)(Y_t - g(\overline{A}_t,\hat\beta)) =0.
\end{equation}
}

An MSM is a semiparametric model
in the sense that it leaves the data generating process unspecified,
subject to the restriction that
the functional $\psi(\overline{a}_t)$ 
has a specific form.
Specifically, let us write
$\psi(\overline{a}_t)$ as
$\psi(\overline{a}_t,p)$ 
to make it clear that
$\psi(\overline{a}_t,p)$ 
depends on the joint density of the data
$p(\overline{a}_T,\overline{y}_T)$
from (\ref{eq::joint}).
The model we are using is then
\begin{equation}\label{eq::calP}
{\cal P} = \Bigl\{ p(\overline{a}_T,\overline{y}_T):\ 
{\rm there\ exists\ }\beta\ {\rm such\ that\ }\psi(\overline{a}_t,p) = g(\overline{a}_t,\beta)\ {\rm for\ all\ }t \Bigr\}.
\end{equation}
The model $g$ is typically chosen to be interpretable.
For example, suppose that
$g(\overline{a}_t,\beta) = \beta_0 + \beta_1 \sum_s a_s$.
Then the effect of the parameter settings is simple
{(i.e., mean outcomes only depend linearly on the amount of cumulative treatment),} 
and the null (of no treatment effect)
simply corresponds to $\beta_1=0$.
It is important to keep in mind that
this is not a model for the entire data generating process,
just for marginal treatment effects, 
i.e., how mean outcomes under different treatment sequences are connected.
\red{Marginal structural models are often
chosen to be some arbitrary but simple parametric model. Instead, we choose to specify 
the marginal structural model $g(a;\beta)$ by the following route:
we tentatively specify a generative model and 
find a
closed form formula $g(a,\beta)$ for $\psi(\overline{a}_t)$.
We then drop the generative model 
and use $g(a,\beta)$ as a MSM.
We explain this in more detail in the next section.}

\begin{remark}
{\it There is a difference between the standard MSM setup and the
one we are considering that warrants mentioning. Typically one
assumes access to $n$ different time series $(Z_1,...,Z_n$), with each
series $Z=\{(A_1,Y_1),...,(A_T,Y_T)\}=(\overline{A}_T,\overline{Y}_T)$
observed for $n$ different independent units (e.g., states).  There,
one could have a different estimating equation at each time, for
example,}
$$ 
\sum_i W_{ti}\ h_t(\overline{A}_{ti})(Y_{ti} - g_t(\overline{A}_{ti},\hat\beta)) =0 
$$ 
{\it where the $i$ subscript denotes weights, treatments, outcomes,
etc.\ for series $i$.  If there are common parameters across
timepoints, then these estimating equations could be combined, for
example by summing over time, or using a generalized method of moments
approach, etc.  However, we model states individually, and so
\textit{do not} assume different states are independent. This leaves
us with one observation per state at each time, which we then combine
across time (but only within state) to obtain estimating equation
\eqref{eq::ee2}.  This represents the trade-off between independence
versus modeling assumptions (e.g., Markov assumptions in the weights,
or linearity in $g(\cdot)$): the less we require of one, the more we
require of the other. }
\end{remark}

\section{Models}
\label{section::models}

Epidemics are
often modeled using
differential equations
that describe the evolution
of certain subgroups over time.
Perhaps the most common is the
SIR (Susceptible, Infected, Recovered) model
(\cite{kermack1927contribution}, \cite{brauer2012mathematical},
\cite{bjornstad2018epidemics})
described by the equations
\begin{align*}
\frac{dS_t}{dt} &= -\frac{\alpha I_tS_t}{N}\\
\frac{dI_t}{dt} &= \frac{\alpha I_tS_t}{N}-\gamma I_t\\
\frac{dR_t}{dt} &= \gamma I_t,
\end{align*}
where $N$ is population size,
$S_t$ is the number of susceptibles,
$I_t$ is the number of infected,
$R_t$ are the removed (by death or recovery) at time $t$
\red{and $\alpha > \gamma$.
Solving the second equation conditional on $S_t$ yields 
$I_t = I_{t-1} e^{\int_{t-1}^t \alpha S_t/N - \gamma dt}$,
which can be discretized as
\begin{equation}
I_t \approx I_{t-1}e^{\alpha S_t/N - \gamma}
\label{eq::SIRsolve}
\end{equation}
when 
$S_u \approx S_t$ for 
all 
$u\in (t-1,t)$.
Without intervention,
the epidemic grows exponentially,
peaks when $S_t/N=\gamma/\alpha$ and then
decays exponentially.}
There are numerous generalizations
of this model
including stochastic versions,
discretized versions and models
with more states besides $S$, $I$ and $R$.

\subsection{The Mobility Model}

Our proposed MSM is
\begin{equation}\label{eq::general}
g(\overline{a}_t,\nu_0,\gimel,f)=
\sum_{s=1}^t f(s,t)e^{\nu_0(s) + \sum_{r=1}^s \gimel(a_r)}
\end{equation}
with nuisance functions
$f$, $\nu_0$ and $\gimel$.
The model is motivated by the SIR model.

The basic idea of the SIR
model is that there is a natural tendency
for an epidemic to increase exponentially at the
beginning.
But there are also elements
that reduce the epidemic
such as the depletion of susceptible
individuals due
to recovery and death.
At the beginning of a pandemic,
reduction of susceptibles
will play a negligible role.
On the other hand,
interventions like lockdowns, school closings etc
can have a drastic effect.
These considerations lead us to
the following \red{working} model.
\red{We use this working model only to
suggest a form for the MSM.}

\begin{figure}
\begin{center}
\begin{tikzpicture}[->, shorten >=2pt,>=stealth, 
node distance=1cm, noname/.style={ ellipse, minimum width=5em, minimum height=3em, draw } ]
\node[] (1) {$A_1$};
\node[circle,fill=red!20,draw] (2) [right= of 1] {$I_1$};
\node (3) [right= of 2] {$Y_1$};
\node (4) [right= of 3] {$A_2$};
\node[circle,fill=red!20,draw] (5) [right= of 4] {$I_2$};
\node (6) [right= of 5] {$Y_2$};
\path (1) edge [thick] node {} (2);
\path (1) edge [thick, bend left=50pt] node {} (4);
\path (1) edge [thick, bend left=50pt] node {} (5);
\path (2) edge [thick] node {} (3);
\path (2) edge [thick, bend right=50pt] node {} (4);
\path (2) edge [thick, bend right=50pt] node {} (5);
\path (2) edge [thick, bend right=50pt] node {} (6);
\path (3) edge [thick] node {} (4);
\path (4) edge [thick] node {} (5);
\path (5) edge [thick] node {} (6);
\end{tikzpicture}
\end{center}
\caption{
\red{
Directed graph illustrating the working model.
Infections $I_t$ are unobserved.
We use this model to find the form 
$g(a;\beta)$ of the causal effect $\psi(a)$.
But when we estimate $\beta$
we use a semiparametric estimating equation approach; we do not fit the 
above model to the data.}}
\label{fig::dag}
\end{figure}
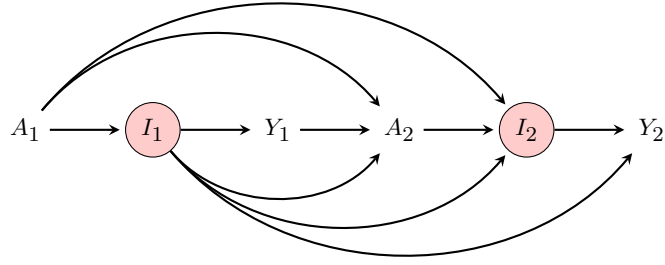

Let $I_t$
denote {\it new} infections in week $t$.
Let
\begin{align}\label{eq::themodel}
A_t &\sim Q_t \nonumber\\
I_t &= I_{t-1}  e^{c_t+\gimel(A_t)} + \delta_t\\
Y_t &= \sum_{s=1}^t f(s,t) I_s + \xi_t \nonumber
\end{align}
where 
$Q_t$ is an arbitrary distribution depending
on $(\overline{A}_{t-1},\overline{I}_{t-1},\overline{Y}_{t-1})$,
$\delta_t$ and
$\xi_t$ are mean 0 random variables (independent of the other variables),
$f(s,t)$ denotes the probability that someone infected at time $s$ dies of COVID at time $t$,
the parameter $c_t$ is a positive number
and $\gimel$ is a smooth function.
\red{Notice that the infection process (second equation) has an exponential growth 
form as in  (\ref{eq::SIRsolve}), but we model 
the exponent directly as a function of mobility and time instead of stipulating a model for the susceptibles $S_t$.}
Here, $c_t$ represents the evolution of the epidemic
without intervention and $\gimel(A_t)$
is the effect of mobility.
We allow $c_t$ to vary with $t$ to make the model more general 
and to allow the spread of Covid-19 to depend on the availability of susceptibles.
We write
\begin{equation}
\label{eq::fst}
f(s,t) = d(s)f_0(s,t)
\end{equation}
where
$d(s)$ is the probability that
someone infected at time $s$ will eventually die 
\red{of COVID}
and 
\red{$f_0(s,t)$}
is the
probability that
someone infected at time $s$ and who will eventually die,
will die at time $t$.
Following \cite{unwin2020state}
we take
$f_0(s,t)$, on the scale of days, to be 
the density of $T_1 + T_2$
where
$T_1$ (time from infection to symptoms) is Gamma with mean 5.1 and coefficient of variation 0.86
and
$T_2$ (time from symptoms to death) is Gamma with mean 18.8 and coefficient of variation 0.45.
The resulting distribution can be accurately
approximated by a Gamma with mean 23.9 days and
coefficient of variation 0.40.
Finally, we integrate this distribution over 7 day bins
to get $f_0(s,t)$ on a weekly scale.
\red{A directed graph illustrating the model
is given in Figure \ref{fig::dag}.}

At this point,
we might use (\ref{eq::themodel}) as our model.
But the $I_t$'s are not observed.
Furthermore,
a non-linear, sequentially specified parametric generative model
can suffer from serious anomalies
when used for causal inference.
In particular,
such a model can suffer from the
{\it null paradox}
(\cite{robins1986new,robins1989analysis,rw}).
This means that there may be no parameter values that
satisfy (i) $Y_t$ is conditionally dependent on past values
of $A_s$ and such that
(ii) the null hypothesis of no treatment effect holds.
We explain this point in more detail
in Section \ref{section::null}.

Instead, 
we apply the $g$-formula
to the model specified by (\ref{eq::themodel})
to find $\E[Y_t^{\overline{a}_t}]$
and use the resulting function
as an MSM.
This yields
\begin{equation}\label{eq::themodel2}
\E[Y_t^{\overline{a}_t}] = \sum_{s=1}^t f(s,t) e^{\nu_0(s) + \sum_{r=1}^s \gimel(A_r)}\equiv
g(\overline{a}_t,\nu_0,\gimel,f)
\end{equation}
where
$\nu_0(s) = \log I_1 +\sum_{r=1}^s c_r$.
(We treat $I_1$ as an unknown parameter that is absorbed into $\nu_0$.)
Now we abandon the working model and just
interpret $g(\overline{a}_t,\nu_0,\gimel,f)$ directly as a model for the
counterfactual $\mathbb{E}[Y^{\overline{a}_t}]$, that is, as an MSM.
Put another way,
we start with the model (\ref{eq::themodel}),
find $g(\overline{a}_t,\nu_0,\gimel,f)=\E[Y^{\overline{a}_t}]$,
and then expand the model to include all joint
distributions that satisfy
$\E[Y_t^{\overline{a}_t}] = g(\overline{a}_t,\nu_0,\gimel,f)$.
This defines the model (\ref{eq::calP}).

The MSM can be fit with the estimating equation (\ref{eq::ee2}), which
corrects for confounding due to past deaths,
not by modeling the entire
conditional process, but by weighting by propensity weights
$W_t$ given by (\ref{eq::Wt}).  This MSM approach allows us to be
agnostic about whether it is our motivating model (\ref{eq::themodel}) that holds, or
some other much more complicated data-generating process.  In fact,
one can go further and take a completely agnostic view, in which the
marginal structural model is not assumed correct at all, but only
viewed as an approximation to the true, and possibly very complex,
underlying counterfactual mean
\citep{neugebauer2007nonparametric}. 

\red{
To summarize, our approach
involves three steps.}

\red{
1. Tentatively specify a working model for infections $I_t$.
}

\red{
2. Find the resulting functional form $g(a;\beta)$ for $\psi(a)$ using the $g$-formula.
We use $g(a;\beta)$ as our MSM.
}

\red{
3. Drop the working model and fit the MSM semiparametrically
without further assumptions on the data generating process.
}

\red{
It is important to emphasize that when we estimate
the causal parameter $\beta$,
we do not assume any model for the epidemic process.
Note that the model for $I_t$
in step 1 is very flexible
but it does assume that the mobility effect is
additive.
An alternative would be to use
a more sophisticated epidemic model
for
$\E[I_t| {\rm past}]$
in step 1.
It would be interesting to do this
and this would help unify the traditional approach
to epidemic modeling with the MSM approach we are using.
However, the implied function
$g(a;\beta)$ would not be in closed form
and it would be very hard to fit this model
especially with only 40 observations.}

\subsection{The Null Paradox}
\label{section::null}

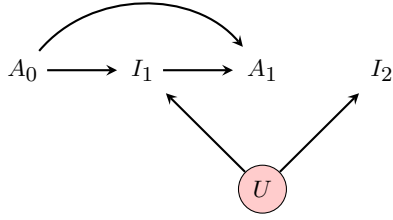
\begin{figure}
\begin{center}
\begin{tikzpicture}[->, shorten >=2pt,>=stealth, 
node distance=1cm, noname/.style={ ellipse, minimum width=5em, minimum height=3em, draw } ]
\node[] (1) {$A_0$};
\node (2) [right= of 1] {$I_1$};
\node (3) [right= of 2] {$A_1$};
\node (4) [right= of 3] {$I_2$};
\node[circle,fill=red!20,draw] (5) [below= of 3] {$U$};
\path (1) edge [thick] node {} (2);
\path (1) edge [thick, bend left=50pt] node {} (3);
\path (2) edge [thick] node {} (3);
\path (5) edge [thick] node {} (2);
\path (5) edge [thick]  node {} (4);
\end{tikzpicture}
\end{center}
\caption{
\red{
The null paradox. The directed graph is a snippet of the time series.
Mobility is $(A_0,A_1)$ and number of infected individuals 
is $(I_1,I_2)$.
The latent variable $U$ is not a confounder as it has no arrows to mobility.
Neither $A_0$ nor $A_1$ have a causal effect on $I_2$.
The variable $I_1$ is a collider, meaning that two arrowheads meet at $I_1$.
This implies that $I_2$ and $(A_0,A_1)$  are dependent conditional on $I_1$.
The estimate of the parameters that relate $I_2$ to $(A_0,A_1)$ in the epidemic model
will be non-zero even though there is no causal effect.}}
\label{fig::nulldag}
\end{figure}

\red{
To see how the null paradox works,
consider a simple example
with four time ordered variables
$(A_0,I_1,A_1,I_2)$
where $A_0$ and $A_1$ are mobility and
$I_1$ and $I_2$ are number of infected,
which we assume are observed.
This is a snippet of the entire time series.
A simple epidemic model is
\begin{align*}
A_0   & \sim p(a_0)\\
\log I_1 &= \beta_0 + \epsilon\\
A_1   & \sim p(a_1|I_1,A_0)\\
\log I_2 &= \theta_0 + \theta_1 A_0 + \theta_2 \log I_1 + \theta_3 A_1 + \delta
\end{align*}
where $\epsilon$ and $\delta$ are, say,  mean 0 Normal random variables.
This is meant to capture exponential growth of $I_t$
(i.e. the SIR model at early times with no recovered individuals).
By applying the $g$-formula,
the causal effect of setting
$A=(A_0,A_1)$ to
$a=(a_0,a_1)$ is
$$
\psi(a)=\E[\log I_2^a] = \theta_0 + \theta_1 a_0 + \theta_2 \beta_0 + \theta_3 a_1.
$$
This means that, if we simulated the epidemic model
with
$A=(A_0,A_1)$ set to
$a=(a_0,a_1)$,
the mean of $\log I_2$ would
precisely be
$\theta_0 + \theta_1 a_0 + \theta_2 \beta_0 + \theta_3 a_1$.
Suppose now that there is an unobserved variable $U$ 
that affects
$I_1$ and $I_2$.
For example, $U$ could represent the general health of the population.
The variable $U$ is not a confounder as it does not affect $A_0$ or $A_1$.
The causal effect is still given by the $g$-formula with no change.
Suppose now that neither $A_0$ or $A_1$ have a causal effect on $I_2$.
The set up is shown in Figure \ref{fig::nulldag}.
Despite the fact that $A_0$ and $A_1$ have no causal effect on $I_2$,
it may be verified that $I_2$ is conditionally dependent on $A_0$ and $A_1$.
(This follows since $I_1$ is a collider on the path $I_2 , U, I_1, A_0, A_1$.)
It follows that
the maximum likelihood estimators $\hat\theta_1$ and $\hat\theta_3$ are not zero
(and in fact converges to a nonzero number in the large sample limit).
The estimated causal effect is
$$
\hat\psi(a)=  \hat\theta_0 + \hat\theta_1 a_0 + \hat\theta_2 \hat\beta_0 + \hat\theta_3 a_1
$$
and will therefore be a function of $a$
even when $a$ has no causal effect.}

\red{
The details of the model
were not important.
A similar model is
\begin{align*}
A_0   & \sim p(a_0)\\
I_1   & \sim p(i_1|A_0)\\
A_1   & \sim p(a_1|I_1,A_0)\\
I_2   & \sim p(i_2|A_0,I_1,A_1)
\end{align*}
where
$\E[I_2|A_0,I_1,A_1] = e^{\beta_0 + \beta_1 A_0 + \beta_2 A_1}I_1$.
In this case
$$
\E[I_2^a] = e^{\beta_0 + \beta_1 a_0 + \beta_2 a_1}\E[I_1|A_0].
$$
The same argument shows that
the estimate will be a function of $a$
even when $a$ has no causal effect.
}

\subsection{Simplified Models}
\label{section::simplified}

The MSM
is not identified without further constraints.
We will take
$\gimel(A_s) = \beta A_s$ so that
\begin{equation*}
\E[Y_t^{\overline{a}_t}] = 
\sum_{s=1}^t f(s,t)e^{\beta \sum_{r=1}^s A_r + \nu_0(s)}.
\end{equation*}
\red{Solving the estimating equation with this model
is unstable and computationally prohibitive.
Hence
we make two approximations.}
First, we take $f_0(s,t)$ in (\ref{eq::fst}) to be a point mass 
at $\delta  = 4$ weeks
(approximately its mean).
Then we get
$$
\E[Y_t^{\overline{a}_t}] =   e^{d(t-\delta) + \nu_0(t-\delta) + \beta M_t}
$$
where
$M_t \equiv M(\overline{a}_t)= \sum_{s=1}^{t-\delta} a_s$.
If we approximate
$\log \mathbb{E}[Y_t^{\overline{a}_t}]$
with
$\mathbb{E}[\log(Y_t^{\overline{a}_t})]$ 
we further obtain
\begin{equation}\label{eq::pointmass}
\mathbb{E}[L_t^{\overline{a}_t}] = \log d(t-\delta) + \nu_0(t-\delta) + \beta M_t
\end{equation}
where $L_t = \log(Y_t+1)$.
Finally,
we take
$$
\nu(t) \equiv \log d(t-\delta) + \nu_0(t-\delta) = \sum_{j=1}^k \beta_j \psi_j(t)
$$
where
$\psi_1,\ldots,\psi_k$
are orthogonal polynomials
starting with $\psi_1(t)=t$.
This model is easy to fit and will be used in 
Section \ref{section::results}.
Note that the probability of dying $d(t)$ is allowed
to change smoothly over time,
which it likely did as hospitals were better prepared during the second wave.
Interestingly, we have consistently found that
using $k=1$ leads to unreasonable results as we discuss in Section 
which means that the disease exponential growth changes with time other than through mobility.
The method for choosing $k$ is described in Section \ref{section::sens}.
Note that
$\partial \E[L_t^{\overline{a}_t}]/\partial a_s = \beta$
for any $s\leq t-\delta$
so $\beta$ has a clear meaning.

The model in (\ref{eq::pointmass}) was used independently in
\cite{shi2020capping} with $k=1$.
They used the model
for curve fitting and they showed that this simple model
fits the data surprisingly well.
However, we find that making $\nu(t)$
non-linear (i.e. $k>1$) is important.

We will also consider
a different approach to fitting the model.
Specifically, we will use deconvolution methods
to estimate the unobserved infection process $I_1,\ldots, I_T$.
The first equation in (\ref{eq::themodel})
implies
$\E[I_t] = e^{\nu(t) + \beta \sum_s A_s}$
suggesting the MSM
$$
\E[L_t^{\overline{a}_t}] = \nu(t) + \beta M_t
$$
which is the same as
(\ref{eq::pointmass})
except that now
$L_t = \log(I_t)$ and
$M_t = \sum_{s=1}^t a_s$
rather than
$M_t = \sum_{s=1}^{t-\delta} a_s$.

\smallskip

\begin{remark}
{\it We have regularized the model by
restricting $\nu(t)$ to have a finite basis expansion.
We also considered a different approach
in which $\nu(t)$ is restricted to be increasing
which seems a natural restriction
if $\nu(t)$ is supposed to represent
the growth of the pandemic in lieu of intervention. 
(This is valid only at the start of the pandemic; later in the pandemic, $\nu$  could be decreasing.)
Using the methods in
\cite{meyer2008inference,meyer2018framework,cgam}
we obtained estimates and standard errors.
The results were very similar to the results
in Section \ref{section::results}.}
\end{remark}

\bigskip

{\bf Counterfactual Estimands.}
Now we discuss some causal quantities that
we can estimate from the model.
Let
$\overline{a}_t = (a_1,\ldots, a_t)$
be a mobility profile of interest.
After fitting the model we will
plot estimates
and confidence 
intervals for counterfactual deaths
\begin{equation}
\label{eq::theta}
\theta_t = \exp\Bigl\{ \E[L^{\overline{a}_t}] \Bigr\}
\end{equation}
under mobility regime $\overline{a}_t$, $t=1, \ldots, T$.

We will consider the following three interventions:
\begin{align*}
{\rm Start\  one\  week\  earlier\ }:\  & \overline a_T = (A_2,A_3,\ldots,, A_{T+1})\\
{\rm Start\  two\  weeks\  earlier\ }:\ & \overline a_T = (A_3, A_4,\ldots, A_{T+2})\\
{\rm Stay\  vigilant\ }:\ & \overline a_T = (A_1,A_2,\ldots,A_9,A_{10}, A_{10}, A_{11}, A_{11}, A_{12}, A_{12}, A_{13}, A_{13},\ldots)
\end{align*}
The first two interventions aim to assess COVID-19 infections if
we had started sheltering in place one and two weeks earlier. The
last intervention halves the slope of the rapid decrease in stay at
home mobility after the initial peak in week 9 that is clearly
visible in Fig.\ref{fig::data}.
See Figure \ref{fig::Interventions}.

\section{Fitting the Model}
\label{section::fitting}

Now we discuss the method
for estimating the model.

\subsection{Fitting the Semiparametric Model}

Recall the MSM
\begin{equation}\label{eq::theMSM}
\mathbb{E}[L_t^{\overline{a}_t}] = \nu(t) + \beta M(\overline{a}_t)
\end{equation}
where $\nu(t) = \sum_{j=1}^k \beta_j \psi_j(t)$.
We estimate $\nu(t)$ and $\beta$
by solving the estimating equation
\begin{equation}\label{eq::thisisit}
\sum_t  h_t(\overline{a}_t) W_t [ L_t - (\hat\nu(t) + \hat\beta M(\overline{a}_t))] =0
\end{equation}
corresponding to (\ref{eq::ee2}). 
We discuss the estimation of the weights $W_t$ in Section \ref{subsection::weights}.
As is often done for MSMs we choose
$$
h_t(\overline{a}_t) =(1,\psi_1(t),\ldots,\psi_k(t),M(\overline{a}_t))^T
$$
since solving the estimating equation
then corresponds to using least squares with weights
$W_t$.
The estimating equation is then the derivative of the weighted sum of squares set to zero.

Recall from (\ref{eq::theta}) that
$\theta_t = e^{\psi(\overline{a}_t)}=e^{\nu(t) + \beta M(\overline{a}_t)}$
which we estimate by
$\hat\theta_t = e^{\hat\nu(t) + \hat\beta M(\overline{a}_t)}$.
We obtain approximate confidence intervals
using the delta method
\red{and the aymptotic normality of estimating equations estimators.
The asymptotic variance is based on
the heteroskedasticity and autocorrelation consistent HAC sandwich estimator
(\cite{Newey}).}

\subsection{Estimating the Stabilized Weights}
\label{subsection::weights}

To estimate the marginal structural model
we need to estimate the stabilized weights
$$
W_t = \prod_{s=1}^t \frac{ \pi(A_s| \overline{A}_{s-1})}{ \pi(A_s| \overline{A}_{s-1}, \overline{Y}_{s-1})};
$$
see (\ref{eq::Wt0}) and (\ref{eq::Wt}).
One approach is to plug in estimates of the 
numerator and denominator densities
into the formula for $W_t$.
But estimating these densities is not easy
and ratios of density estimates can be unstable.
The problem is exacerbated
when we multiply densities.
Instead we use a moment-based approach as in
\cite{fong2018covariate,zhou2018residual}.
The idea is to estimate the vector of weights
$W_1,\ldots, W_T$
by noting that they need to satisfy certain moment constraints.
Our method is similar to
the approach in
\cite{zhou2018residual}.

We rewrite
$W_t = \prod_{s=1}^t V_s$ where
$$
V_s \equiv V_s(\overline{A}_s,\overline{Y}_{s-1})=
\frac{\pi(A_s|\overline{A}_{s-1})}{\pi(A_s|\overline{A}_{s-1},\overline{Y}_{s-1})}.
$$
Let
$\tilde h_1(a_t)$ and
$\tilde h_2(y_{t-1})$ 
be arbitrary functions and define their centered versions by 
\begin{align*}
h_1(a_t) &= \tilde h_1(a_t) - \mu_t\\
h_2(y_{t-1}) &= \tilde h_2(y_{t-1}) - \nu_t
\end{align*}
where the conditional means are
\begin{align*}
\mu_t &\equiv \mu_t(\overline{A}_{t-1})=\E[\tilde h_1(A_t) | \overline{A}_{t-1}]\\
\nu_t &\equiv \nu_t(\overline{A}_{t-\delta-1},\overline{Y}_{t-2})=\E[\tilde h_2(Y_{t-1}) | \overline{A}_{t-\delta-1},\overline{Y}_{t-2}].
\end{align*}
Weighted products of these functions have mean zero since
\begin{align*}
\E[ h_1(A_t) h_2(Y_{t-1}) W_t ] &=
\int \cdots \int
h_1(a_t) h_2(y_{t-1}) p(\overline{a}_t,\overline{y}_{t-1}) 
W_t (\overline{a}_t,\overline{y}_{t-1})\,
d\overline{a}_t\, d \overline{y}_{t-1}\\
&\hspace{-1in}=
\int \cdots \int
h_1(a_t) h_2(y_{t-1}) 
\pi(a_t|\overline{a}_{t-1},\overline{y}_{t-1}) p(y_{t-1}|\overline{a}_{t-1},\overline{y}_{t-2})
p(\overline{a}_{t-1},\overline{y}_{t-2})\\
&\ \ \ \ \ \ \times
\frac{\pi(a_t|\overline{a}_{t-1})}{\pi(a_t|\overline{a}_{t-1},\overline{y}_{t-1})}
\left(\prod_{s=1}^{t-1}V_s\right)\ \ 
d\overline{a}_t\, d \overline{y}_{t-1}\\
&\hspace{-1in}=
\int \Biggl\{ \omega(\overline{y}_{t-2},\overline{a}_{t-1})
\int h_1(a_t) \pi(a_t|\overline{a}_{t-1}) d a_t\ 
\int h_2(y_{t-1}) p(y_{t-1}|\overline{a}_{t-1},\overline{y}_{t-2}) dy_{t-1}
\Biggr\} d \overline{a}_{t-1}\, d \overline{y}_{t-2}\\
&\hspace{-1in} = 0
\end{align*}
from the definition of $h_1$ and $h_2$,
where
$$
\omega(\overline{y}_{t-2},\overline{a}_{t-1}) =
p(\overline{y}_{t-2},\overline{a}_{t-1})\prod_{s=1}^{t-1} V_s.
$$
Thus, the weights are characterized by the moment constraints
\begin{equation}\label{eq::ww}
\E[ h_1(A_t) h_2(Y_{t-1}) W_t ] =0.
\end{equation}

As in 
\cite{zhou2018residual}
we estimate the weights
by finding $W_t$
to satisfy
$\E[ h_1(A_t) h_2(Y_{t-1}) W_t ] =0$
for a set of functions
$h_1, h_2$.
This requires estimating these moments
and estimating $\mu_t$ and $\nu_t$.
To proceed, we make a Markov assumption, namely
$$
\E[\tilde h_1(A_t) | \overline{A}_{t-1}] = \E[\tilde h_1(A_t) | A_{t-1},\ldots, A_{t-k}]
$$
and
$$
\E[\tilde h_2(Y_{t-1}) | \overline{A}_{t-\delta-1},\overline{Y}_{t-2}] = 
\E[\tilde h_2(Y_{t-1}) | A_{t-1-\delta},\ldots, A_{t-k-\delta},{Y}_{t-2},\ldots, Y_{t-k}]
$$
for some $k$.
We will use $k=1$ in our analysis.
Moreover, we assume homogeneity so that
the functions
$\mu_t$ and $\nu_t$ do not depend on $t$.
Under the homogeneous Markov assumption,
$\mu$ and $\nu$ can be
estimated by regression.
For example, if $k=1$,
$\mu$ can be estimated by regressing
$\tilde{h}_1(A_2),\dots,\tilde{h}_1(A_T)$ on
$A_1,\ldots, A_{T-1}$.
(We tried both linear and nonparametric regression
and obtained similar weights from each approach
so we have used linear regression in our results.)
The sample versions of the moment conditions
(\ref{eq::ww})
are then
$$
\frac{1}{T}\sum_t H_{tj} W_t = 0
$$
where
$$
H_{tj} =
 (\tilde{h}_{1j}(A_t)-\hat \mu_j)(\tilde h_{2j}(Y_{t-1})-\hat\nu_j) 
$$
and
$\{( \tilde{h}_{1j},\tilde{h}_{2j}):\ j=1,\ldots, J\}$
are a set of pairs of functions,
$\hat\mu_j$ is the estimate of
$\E[\tilde h_1(A_t) | A_{t-1},\ldots, A_{t-k}]$ and
$\hat\nu_j$ is the estimate of
$\E[\tilde h_2(Y_{t-1}) | A_{t-1-\delta},\ldots, A_{t-k-\delta},{Y}_{t-2},\ldots, Y_{t-k}]$.

The moment conditions do not completely specify the weights.
As in the above references
we add a regularization term, in this case,
$(1/2)\sum_t (W_t-1)^2$
and we require
$\sum_t W_t = T$.
This leads to the following minimization problem:
minimize $W_1,\ldots, W_T$ in
\begin{equation}
\frac{1}{2}\sum_t (1-W_t)^2 + 
\lambda_0 \sum_t (W_t - T) + \sum_{j=1}^J \lambda_j \sum_t W_t H_{tj}
\end{equation}
where
the $\lambda_j$'s are Lagrange multipliers.
The solution to the minimization is
\begin{equation}\label{eq::W}
W = \mathbf{1} - H (H^T H)^{-1} [ H^T \bf{1}-D]
\end{equation}
where
$W = (W_1,\ldots, W_T)$,
$\mathbf{1}$ is a vector of $1's$,
$\mathbf{D} = (T,0,\ldots,0)^T$ and
$$
H =
\left(
\begin{array}{cccc}
1      & H_{11} & \cdots & H_{1N}\\
1      & H_{21} & \cdots & H_{2N}\\
\vdots & \vdots & \vdots & \vdots\\
1      & H_{T1} & \cdots & H_{TN}
\end{array}
\right)
$$
and $N$ is the total number of moment constraints.
In our case we choose
$h_{11}(a)=a$,
$h_{12}(a)=a^2$,
$h_{21}(y)=y$,
$h_{22}(y)=y^2$.

To include other
time varying confounders $X_t$
one should replace $h_2(y_{t-1})$ with two functions:
$$
h_2(y_{t-1})=
\tilde{h}_2(y_{t-1}) - \E[\tilde{h}_2(y_{t-1}) | \overline{X}_{t-1},\overline{A}_{t-1},\overline{Y}_{t-2}]
$$
and
$$
h_3(x_{t-1})=
\tilde{h}_3(x_{t-1}) - \E[\tilde{h}_3(x_{t-1}) | \overline{X}_{t-2},\overline{A}_{t-1},\overline{Y}_{t-2}].
$$

The steps
for fitting the model
are summarized in Fig.(\ref{fig::alg}).
\red{
Note that we cannot include past infections
as a confounder since this variable is not observed. 
We choose not to include past cases or hospitalizations because the 
former is terribly biased downward at the beginning of the epidemic, and reliable data for the second is difficult to obtain.
We need to assume that
adjusting for past deaths serves
as an adequate surrogate for infections, cases and hospitalizations.
We address the more general problem of unoberved
confounding in Section 6.2.}

\begin{figure}[t!]
\fbox{\parbox{6in}{
\begin{center}
\begin{enumerate}
\item Choose the order $k$ of the Markov assumption.
\item Choose $J$ pairs of functions
$\Bigl\{(\tilde h_{1j}(a),\tilde h_{1j}(y)):\ j=1,\ldots J\Bigl\}$.
\item Estimate
$\mu_j=\E[\tilde h_{1j}(A_t) |A_{t-k},\ldots, A_{t-1}]$
and
$\nu_j=\E[\tilde h_{2j}(Y_{t-1})|A_{t-k-\delta-1},\ldots, A_{t-\delta-1},
Y_{t-1-k},\ldots, Y_{t-2}]$
by regression.
\item Compute the weights $W_1,\ldots, W_n$ from $(\ref{eq::W})$.
\item Fit the model
$L_t = \beta \sum_{i=1}^{t-\delta} A_s + \nu(t) + \epsilon_t$
using weighed least squares
with weights
$W_1,\ldots, W_n$.
\end{enumerate}
\end{center}
}}
\caption{Steps for fitting the model.}
\label{fig::alg}
\end{figure}

\section{Results}
\label{section::results}

In this section we give results
for the mobility measure `proportion of people staying at home.'
We begin by showing the results
of fitting the MSM to each state.
Then we report on various types of sensitivity analysis.

\subsection{Main Results}

\begin{figure}
	\centering
\begin{subfigure}[t]{.5\textwidth}
	\centering
	\includegraphics[width=.9\linewidth]{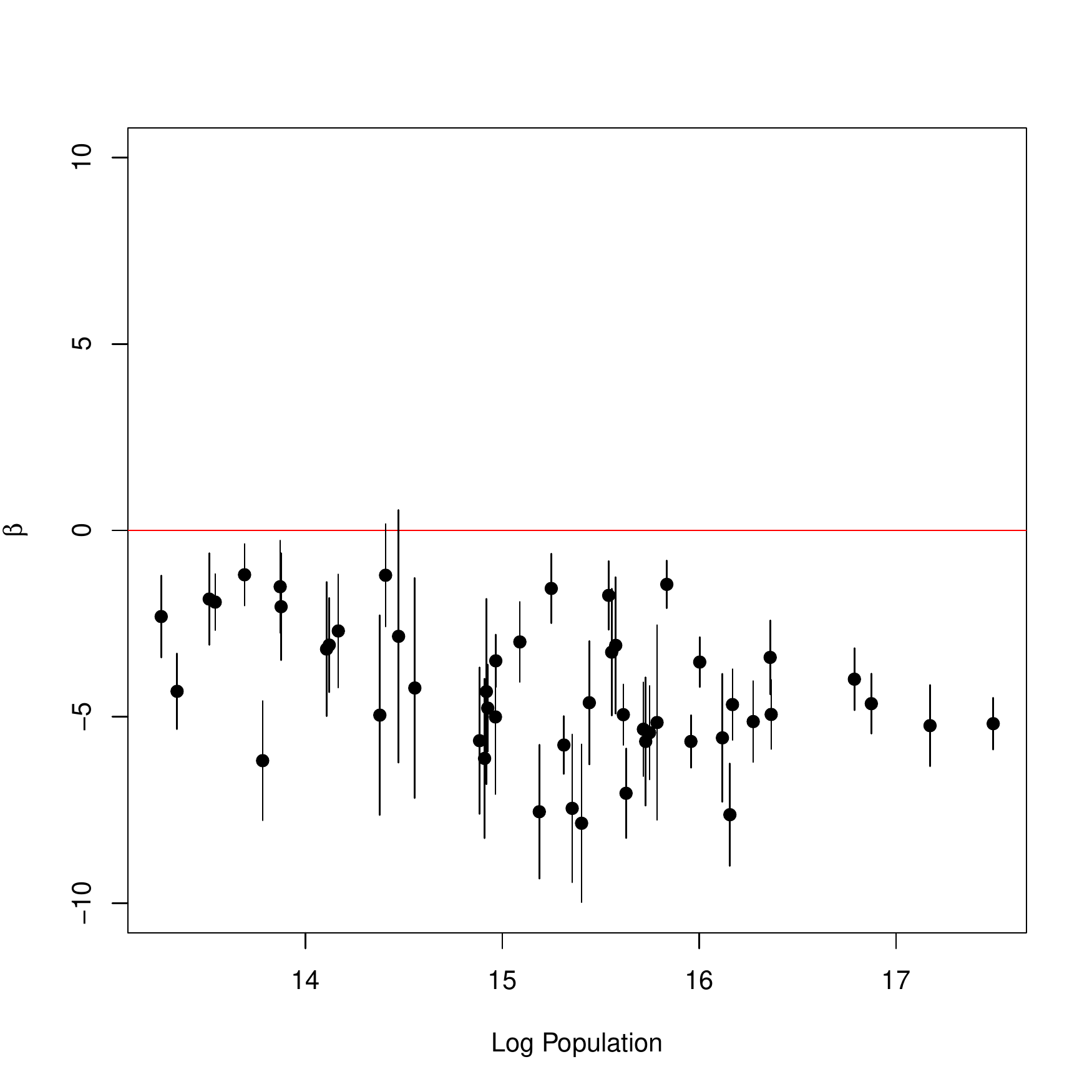}
	\caption{Plot of $\hat\beta$ and 95\% confidence interval
		from the marginal structural model (\ref{eq::theMSM}) for each state, versus state log population. 
		A value of $\beta= -5$, for example, means that log deaths are reduced by 5 if $A_s$ is increased by
		one percent at any time $s$.}
	\label{fig::Plot_MSM_Beta}
\end{subfigure}%
\begin{subfigure}[t]{.5\textwidth}
	\centering
	\includegraphics[width=.9\linewidth]{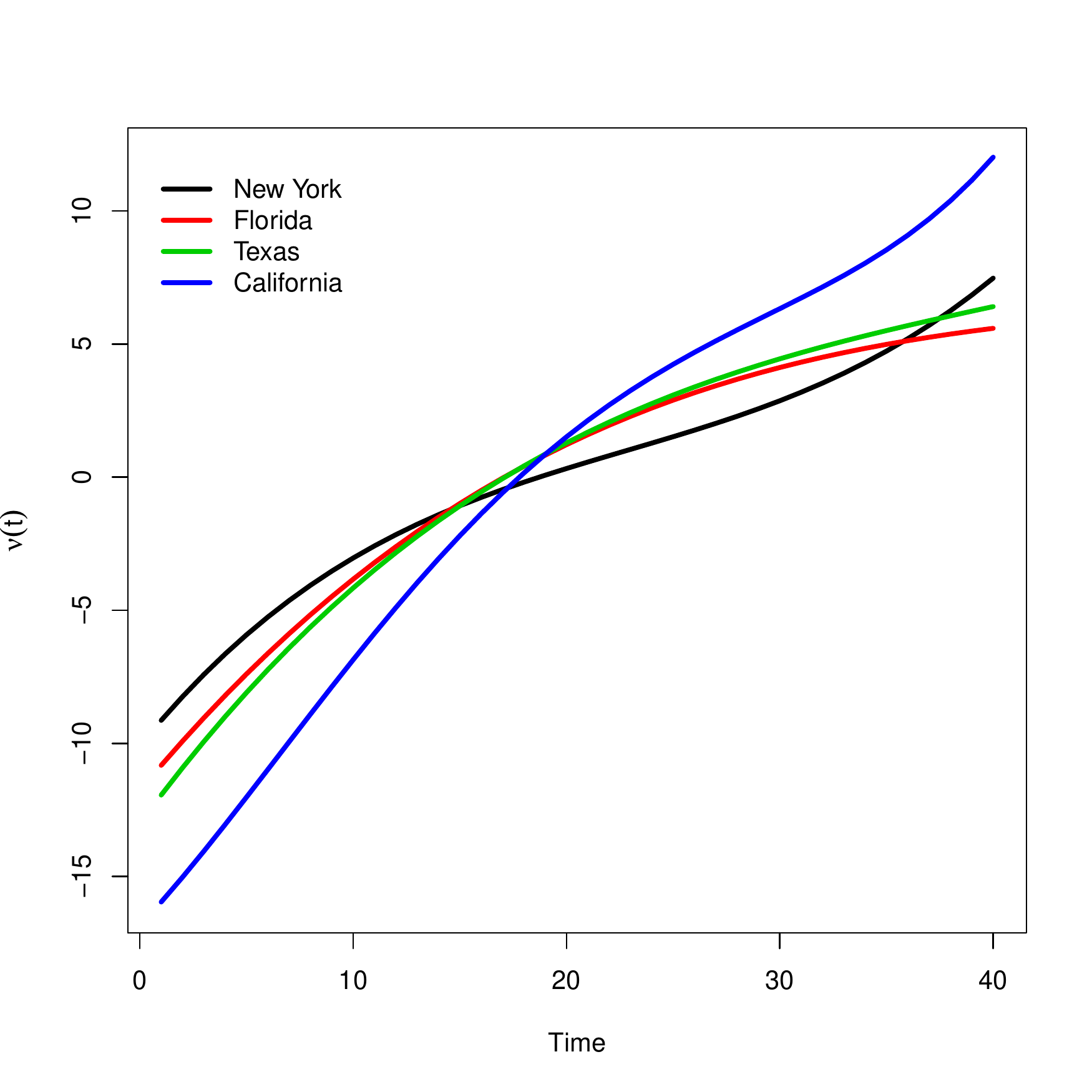}
	\caption{Plot of $\hat\nu(t)$ for four populous states.}
	\label{fig::smooth}
\end{subfigure}
\caption{Estimates of the MSM parameters defined in \eqref{eq::theMSM}.}
\end{figure}

Figure \ref{fig::Plot_MSM_Beta} shows
95 percent confidence intervals for $\hat\beta$ for each state
from the marginal structural model in (\ref{eq::theMSM}). 
We computed standard errors as if the weights were known, which results in valid
but potentially conservative inference as long as the weight models are
correctly specified \citep{tsiatis2007semiparametric}.
The estimates are mostly negative, 
as would be expected, since 
higher $A_s$ means less mobility.
Interestingly,
we find that there turns out to be little
confounding due to past deaths, 
as the fits with and without the estimated weights (not shown)
are very similar.
Nevertheless, we 
keep the weights in all the fits
as a safeguard.
In Section \ref{section::sens}
we investigate this further by doing a sensitivity analysis.

\red{Figure \ref{fig::smooth} shows the estimated
smooth function $\hat\nu(t)$ in (\ref{eq::theMSM})
for four states.
The functions are increasing with slopes tapering off as time goes by, and picking back up again
in NY and CA around week 35, consistent with deaths rising at that time in these two states; see
Figure \ref{fig::data}.
The shape of $\hat\nu(t)$ is consistent with the usual epidemic dynamics where it is assumed that 
this component should initially grow (linearly with no interventions and with 
an infinite pool of susceptibles) on the log-scale at the start of the epidemic 
and then decrease.
Some of the non-linearity probably reflects the fact that
the probability $d(t)$ of dying decreases over time due to better hospital treatment,
social distancing changes, and the number of susceptibles to COVID-19 
decreases over time as recovered patients are likely immune for some period post-infection.
}

\begin{figure}
\begin{center}
\includegraphics[width=5.5in]{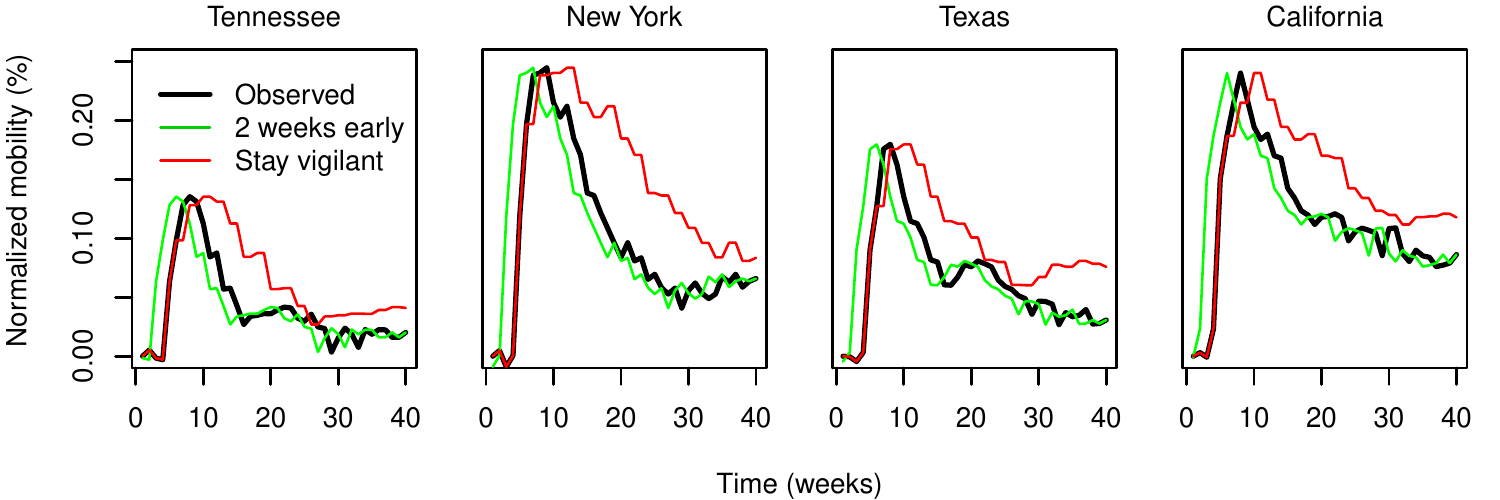}
\end{center}
\vspace{-.15in}
\caption{The observed mobility curves and hypothetical interventions for four states.
Mobility has been standardized to have value 0 at the beginning of the series. All plots are on the same scale.}
\label{fig::Interventions}
\end{figure}

\begin{figure}
\begin{center}
\includegraphics[scale=0.8]{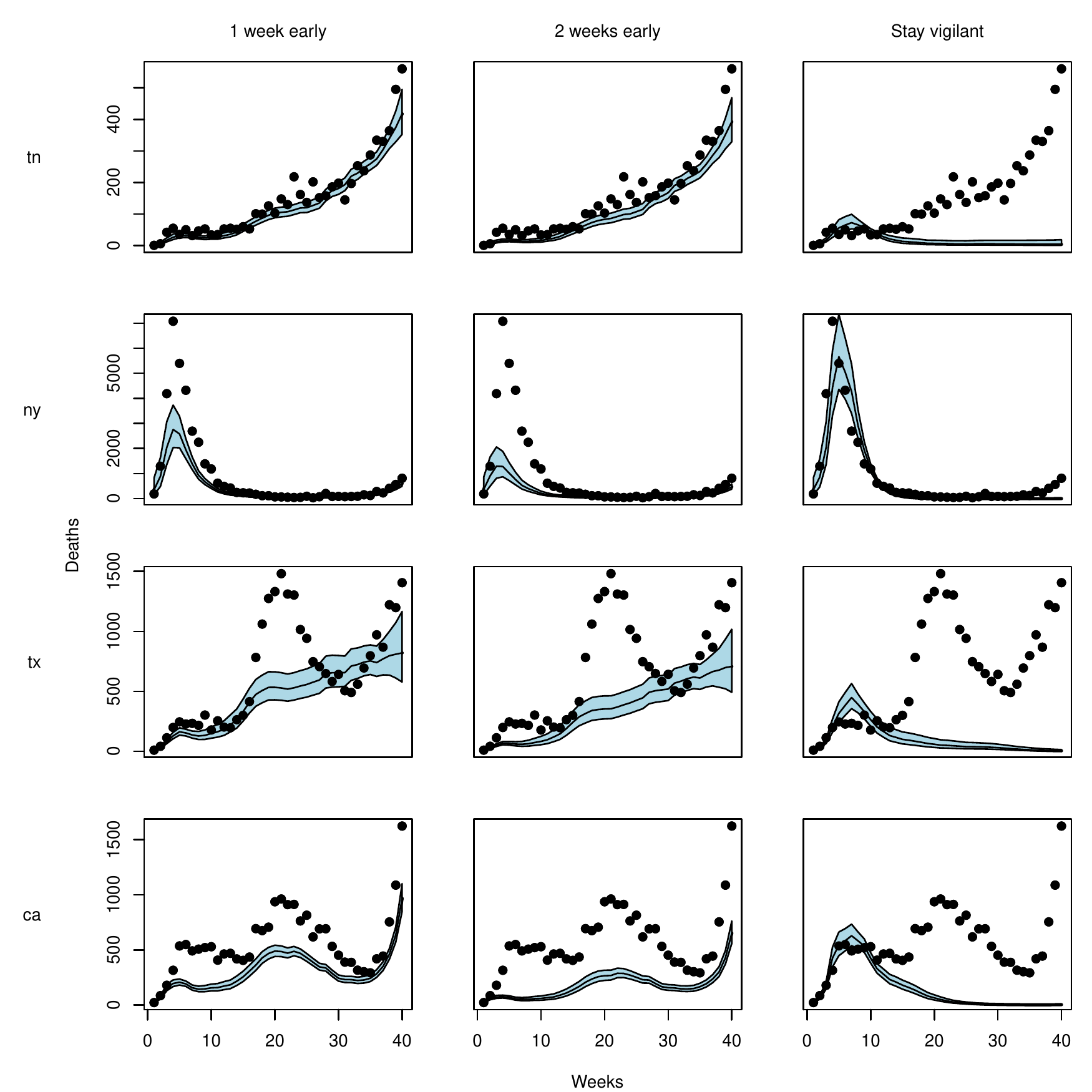}
\end{center}
\vspace{-.15in}
\caption{Pointwise 95\% confidence bands for 
deaths $\theta_t = \exp(\E [L^{\overline{a}_T}])$
for the three mobility scenarios $\overline{a}_T$ 
described at the end of section 4; see also Figure \ref{fig::Interventions}. 
Each row is a different state.
Each column is a different scenario,
start one week early,
start two weeks early
and stay vigilant. The epidemic in NY started early so
staying at home sooner had a large impact. The same is true for
PA, IL, MI, NJ, MA. Staying home earlier would not have had as much
impact in states such as TN that did not suffer the epidemic
early. Staying more vigilant would have had a large impact except for New York.
Some lack of fit in the early time period is evident in Texas, where counterfactual deaths
exceed observed deaths under `stay vigilant'
where mobility has not yet been changed.}
\label{fig::theta}
\end{figure}

Next we consider counterfactual deaths $\theta_t=\exp(\E[L^{\overline{a}_T}])$
in (\ref{eq::theta}) for the three mobility scenarios 
described at the end of section 4; two mobility scenarios are shown in Figure
\ref{fig::Interventions} for four states.
Figure \ref{fig::theta} shows the estimates and pointwise 95 percent
confidence bands for $\theta_t$ for these four states. The plots for all states are in the Supplement.

\begin{figure}
\begin{center}
\includegraphics[width=5.5in]{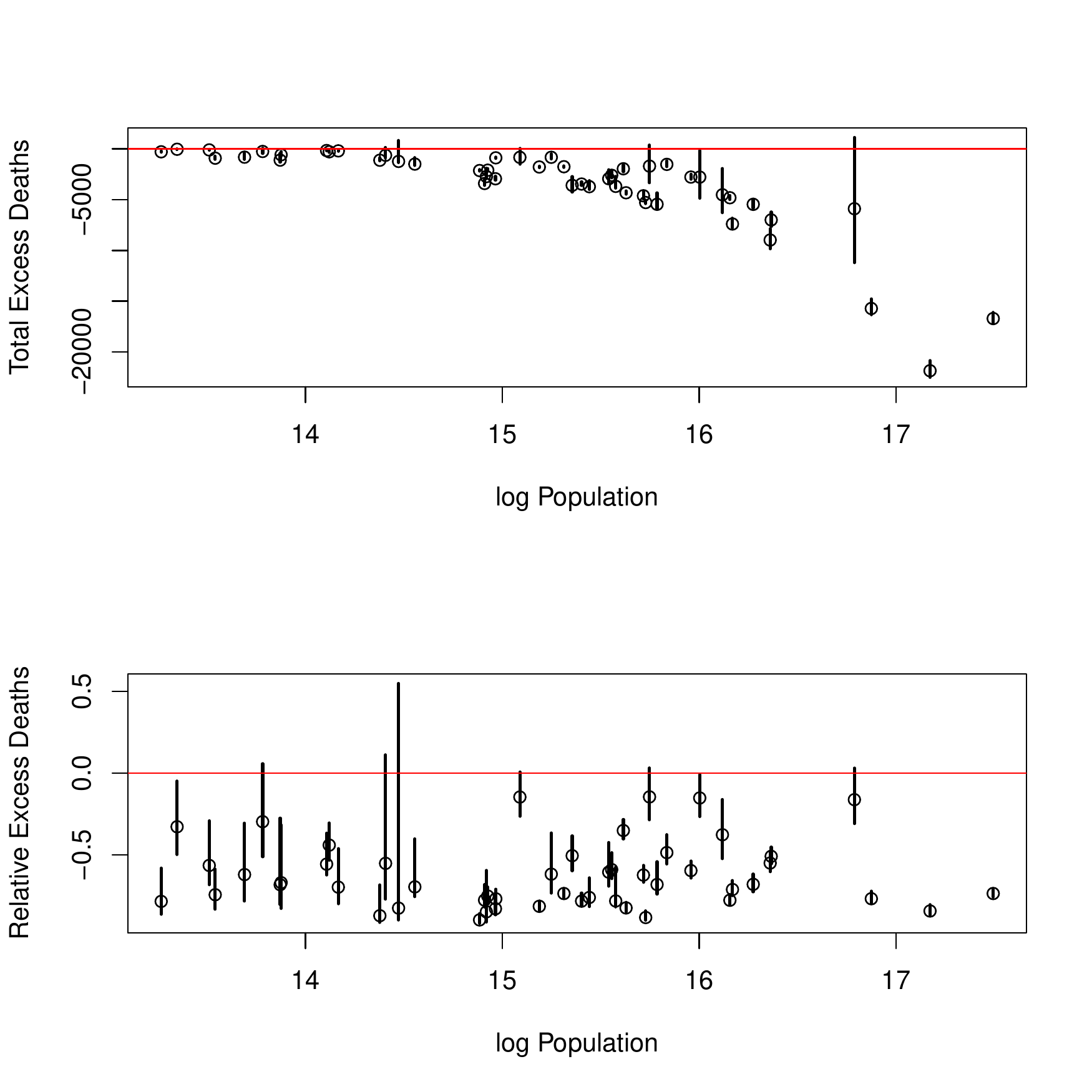}
\end{center}
\vspace{-.15in}
\caption{95\% confidence intervals for total excess deaths
$\sum_t \exp\left( \mathbb{E}[L^{\overline{a}_t}]\right) - \sum_t
Y_t$ (top) and relative excess deaths $\left(\sum_t \exp\left(
\mathbb{E}[L^{\overline{a}_t}]\right) - \sum_t Y_t \right) / \sum_t Y_t$ 
(bottom) under the `stay vigilant' scenario. 
The confidence intervals for NY (fourth from right) and a handful of
other states include zero and suggests that staying more vigilant
would not have significantly impacted the death toll. On the other
hand, many states, small and large, could have reduced their death
tolls by over a half.}
\label{fig::total}
\end{figure}

Finally, Figure \ref{fig::total} shows
95 percent confidence intervals for
$\sum_t \exp\left( \mathbb{E}[L^{\overline{a}_t}]\right) - \sum_t Y_t$ and
for $\left(\sum_t \exp\left( \mathbb{E}[L^{\overline{a}_t}]\right) - \sum_t Y_t \right) / \sum_t Y_t$
under the `stay vigilant' scenario.
We refer to these as total and relative excess deaths,
where a negative excess means that lives would be saved.
Of course, this number is larger for more populous states, although relative to the total
number of observed deaths, all states small and large would have benefited equally from 
more sustained vigilance. Note that the confidence interval for New York (fourth from right) is very large.
New York experienced the pandemic early and responded with
large values of $A_s$
so it is believable that further vigilance may not
have a large effect.

\red{
We now compare our results to those in 
\cite{unwin2020state}.
They use a sophisticated model
of the epidemic dynamics so
a direct comparison is difficult.
They estimate a parameter
$R_t$ that measures how many individuals an infected person will infect.
Using a Bayesian approach,
they find a 95 percent posterior interval
for the change in $R_t$
for the U.S. when setting mobility to its maximum value 
is [26.5,77.0].
The log of the change in $R_t$
is roughly equivalent to $-\beta$ in our setting.
On the log scale, their interval is
[3.3,4.3].
Our effect sizes are similar and slightly larger for the large states.
For the middle sized states our effect estimates vary somewhat 
and are sometimes larger and sometimes smaller than theirs.
Overall,
the effect estimates
are quite similar which is reassuring
given how vastly different the methods are.
Another point of comparison is \cite{chernozhukov2020causal}
who consider a very ambitious model
which includes multiple policy interventions and
multiple mobility measures (which they call behavior) simultaneously
and the model is over all states.
Their estimate of the mobility effect on log cases is
-0.54 with a standard error of .19.
Unlike \cite{unwin2020state},
this estimate is very different from ours.
We do not know why the effect size is so different from ours.
They are using a different measure of mobility 
(they used Google mobility)
which might have some effect.
It is possible that some of the mobility effect might be absorbed
into their policy effect which could happen if there is model misspecification.}

\subsection{Sensitivity Analysis}
\label{section::sens}

We have made a number of strong
assumptions in our model.
Our preference would be to weaken these assumptions
and use nonparametric methods
but the data are too limited to do so.
Instead, we now assess the sensitivity
of the results to various assumptions.
We consider various perturbations of our analysis.
These include:
(1) changing the model/estimation method
(we replace the MSM with a generative model),
(2) assessing the Markov assumption
(which was used to estimate the weights),
(3) checking the accuracy of the point mass approximation
(which was used in Section \ref{section::simplified}
to simplify the model) and
(4) assessing sensitivity to unmeasured confounding
(we have assumed that the only time varying
confounders are past values of mobility and death).

\bigskip

\noindent
{\it 1. An Alternative Model.}
Here we compare 
the results from the MSM 
in (\ref{eq::theMSM})
to the time series AR(1) model:
\begin{equation}\label{eq::ar1}
L_t = L_{t-1} + \beta A_{t-\delta} + r(t) + \epsilon_t
\end{equation}
where
$r(t)$ is a polynomial of degree $k-1$.
This says that, apart from random error,
$L_t$ differs from $L_{t-1}$
for two reasons, mobility $A_{t-\delta}$
and the natural increase $r(t)$ due to epidemic
dynamics (at the start of the epidemic).
If we apply the $g$-formula  in (\ref{eq::the-g}) to this model,
we find
$\E[L_t^{\overline{a}_t}] = \beta M(\overline{a}_t) + \nu(t)$
where $\nu(t) = \sum_{s=1}^t r(s)$
is a polynomial of order $k$.
Hence, this model is consistent with the MSM.
In other words, this model
is contained in the semiparametric model 
${\cal P}$ defined in (\ref{eq::calP}). 
This model resembles Robins' {\it blip models}
(\cite{robins2000marginal2,vansteelandt2014structural})
as it measures the effect of one blip of treatment $A_{t - \delta}$
so we will refer to (\ref{eq::ar1}) as the blip model.
We will fit (\ref{eq::ar1}) by least squares.
There are three reasons for fitting this model.
First, it as a point of comparison for the MSM.
Second, 
we are able to check residuals and model fit.
Third, since it is a regression model,
we can use AIC to choose the degree $k-1$
of $r(t)$.
We also use this choice of $k$ in the MSM.
The degree $k$ chosen by AIC is typically $k=1$ for small states
and $k=3$ or $k=4$ for the larger states.
A plot of the selected degree versus log population and versus
log deaths is in the supplementary material.

\begin{figure}
\begin{center}
\includegraphics[scale=0.6]{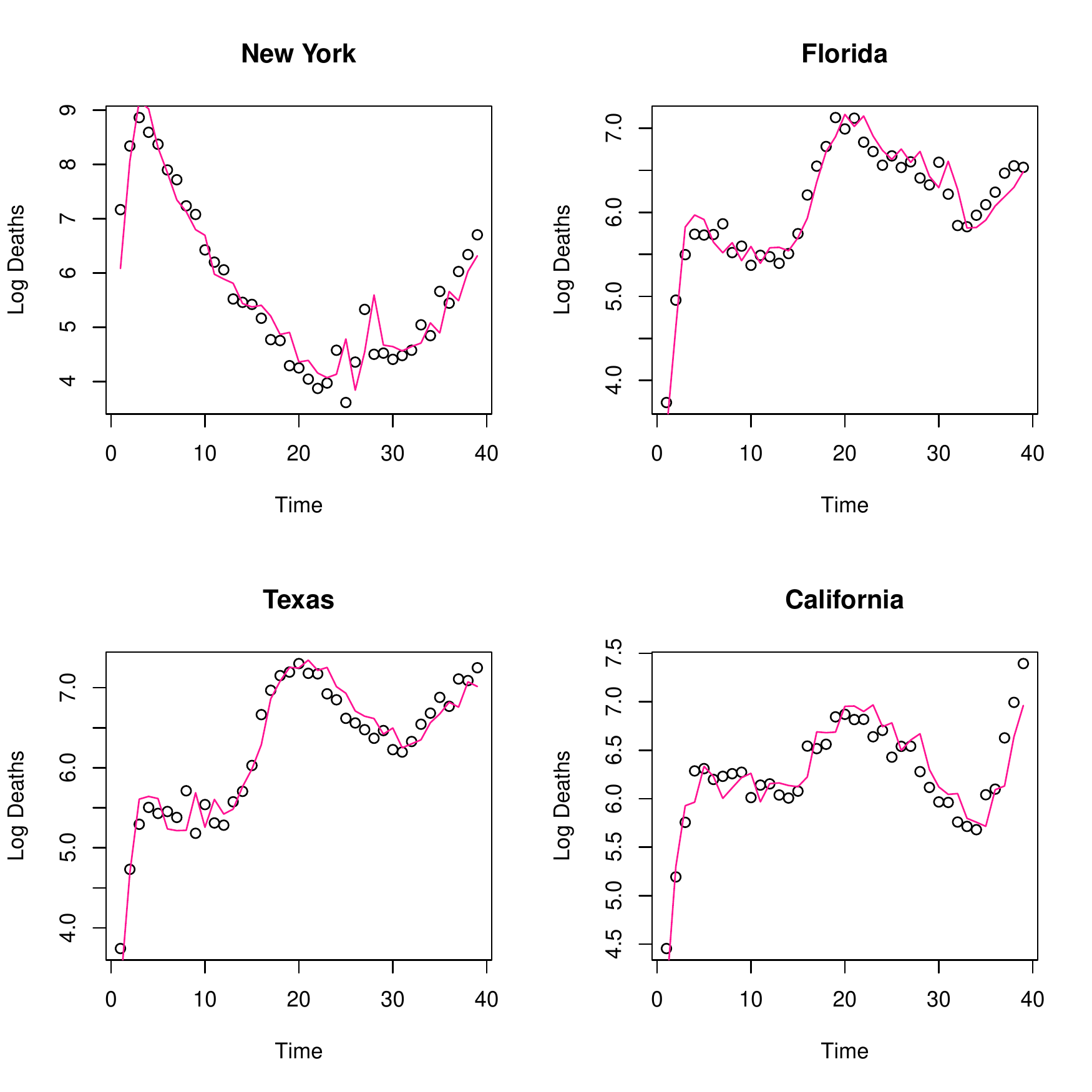}
\end{center}
\vspace{-0.2in}
\caption{Observed log deaths in four states as functions of time with estimates (red) from the blip model in (\ref{eq::ar1}).}
\label{fig::blip_fits}
\end{figure}

The left plot in Figure \ref{fig::blip} shows the
estimates of $\beta$ and 95 percent confidence intervals for all the states from the blip model in (\ref{eq::ar1}),
and the right plot compares the estimates of $\beta$ from the MSM and blip models,
where we see the similarity of the inferences.
Since the blip model is a regression model, it makes sense to compare
the observed data to the fits.
Fig \ref{fig::blip_fits} shows the fitted values and the data
for four states.
The fit is not perfect but is reasonable. 
There are some large outliers in some states,
mostly in the first few weeks of the
pandemic where mobility $A_t$ and log deaths $L_t$ change rapidly.
Because of this we also fitted a robust
regression but the results did not change much.

\begin{figure}
	\begin{subfigure}[t]{.5\textwidth}
		\centering
		\includegraphics[width=.9\linewidth]{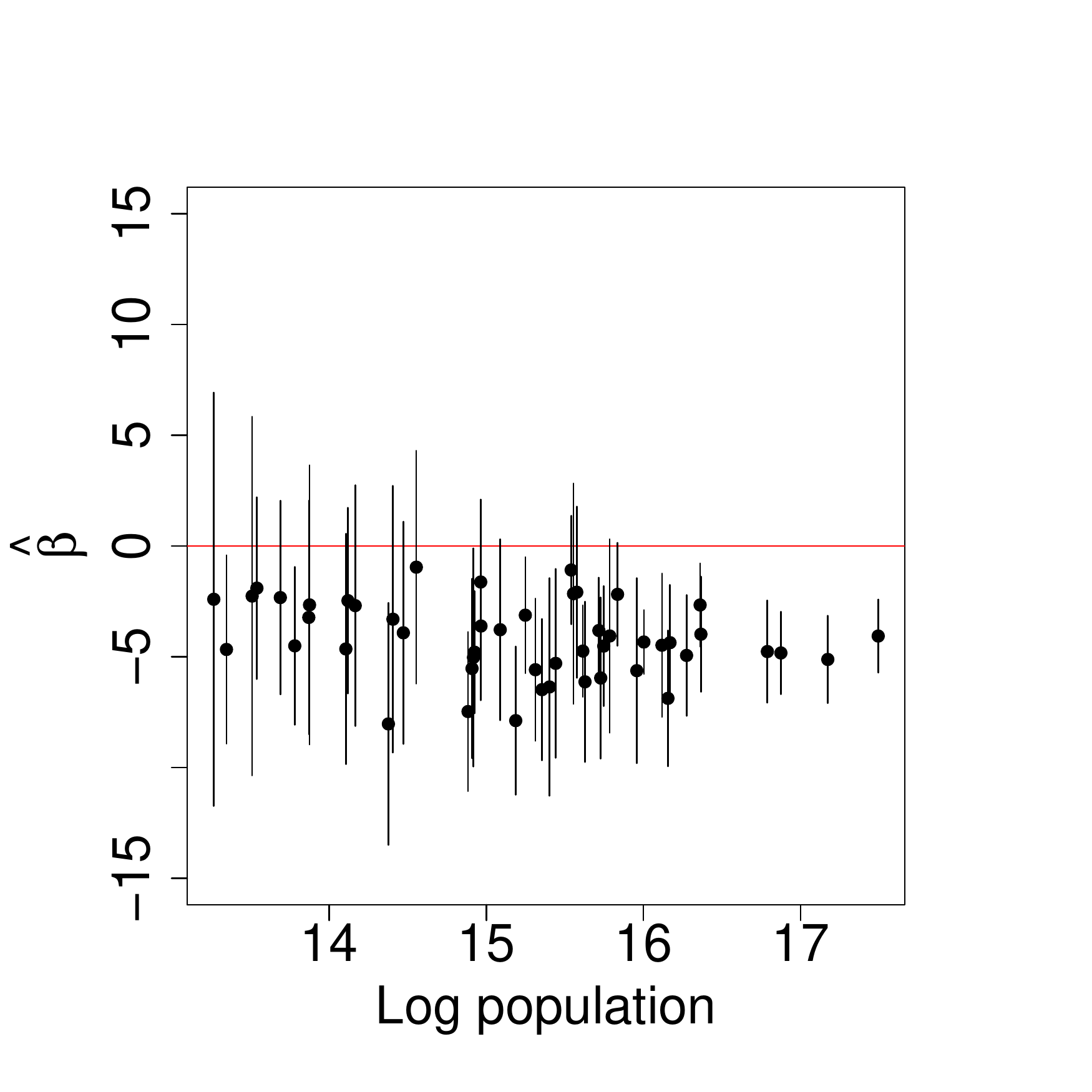}
		\caption{Estimates of $\beta$ from the blip model in \\ (\ref{eq::ar1}) with 95\% confidence intervals. }
	\end{subfigure}%
	\begin{subfigure}[t]{.5\textwidth}
		\centering
		\includegraphics[width=.9\linewidth]{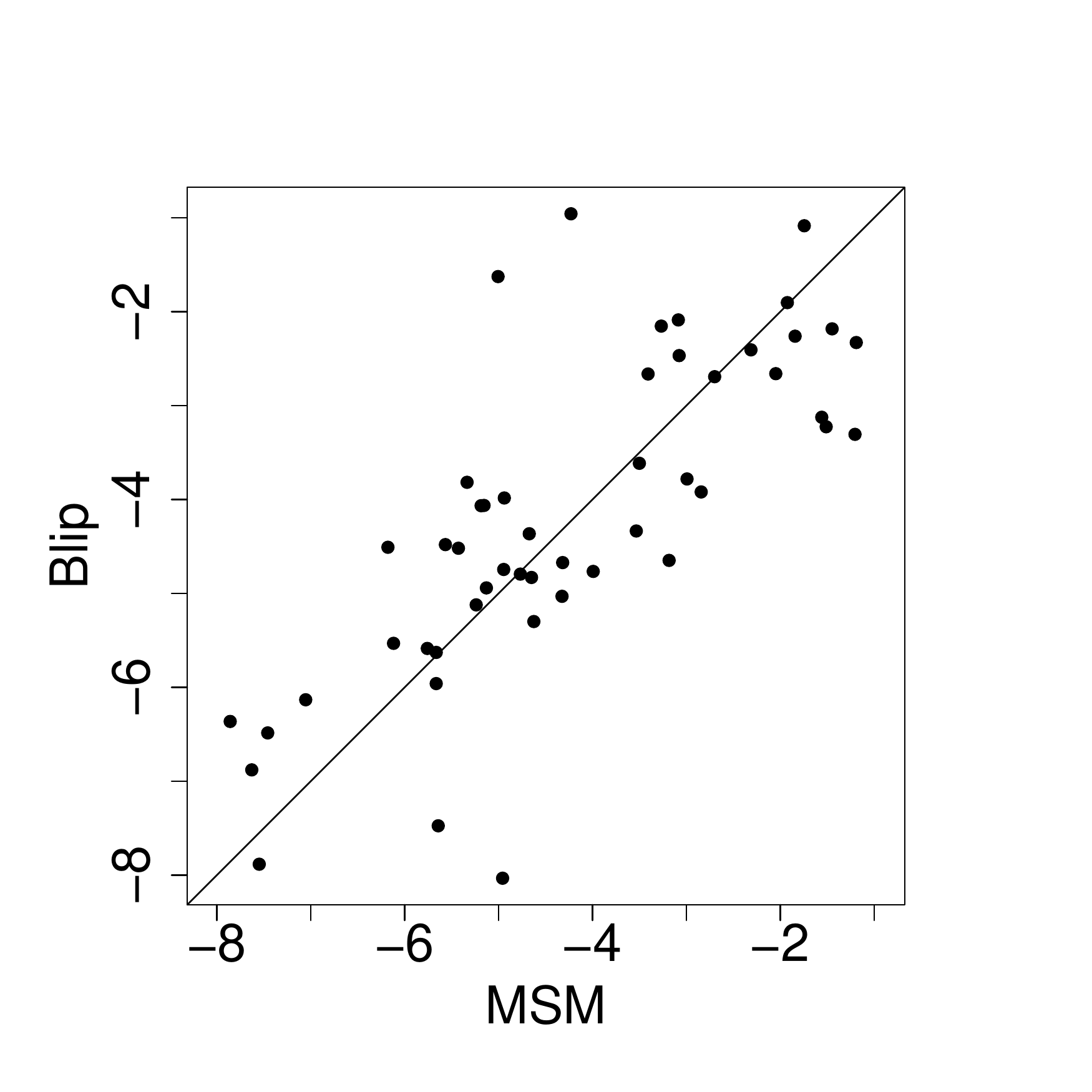}
		\caption{Comparison of estimates of $\beta$ from the blip model and the MSM in (\ref{eq::theMSM}).}
	\end{subfigure}
\caption{Estimates from the blip model compared with estimates from the MSM model.\label{fig::blip}}

\end{figure}
\vspace{0.2in}

\bigskip

\begin{figure}
\begin{center}
\includegraphics[scale=0.75]{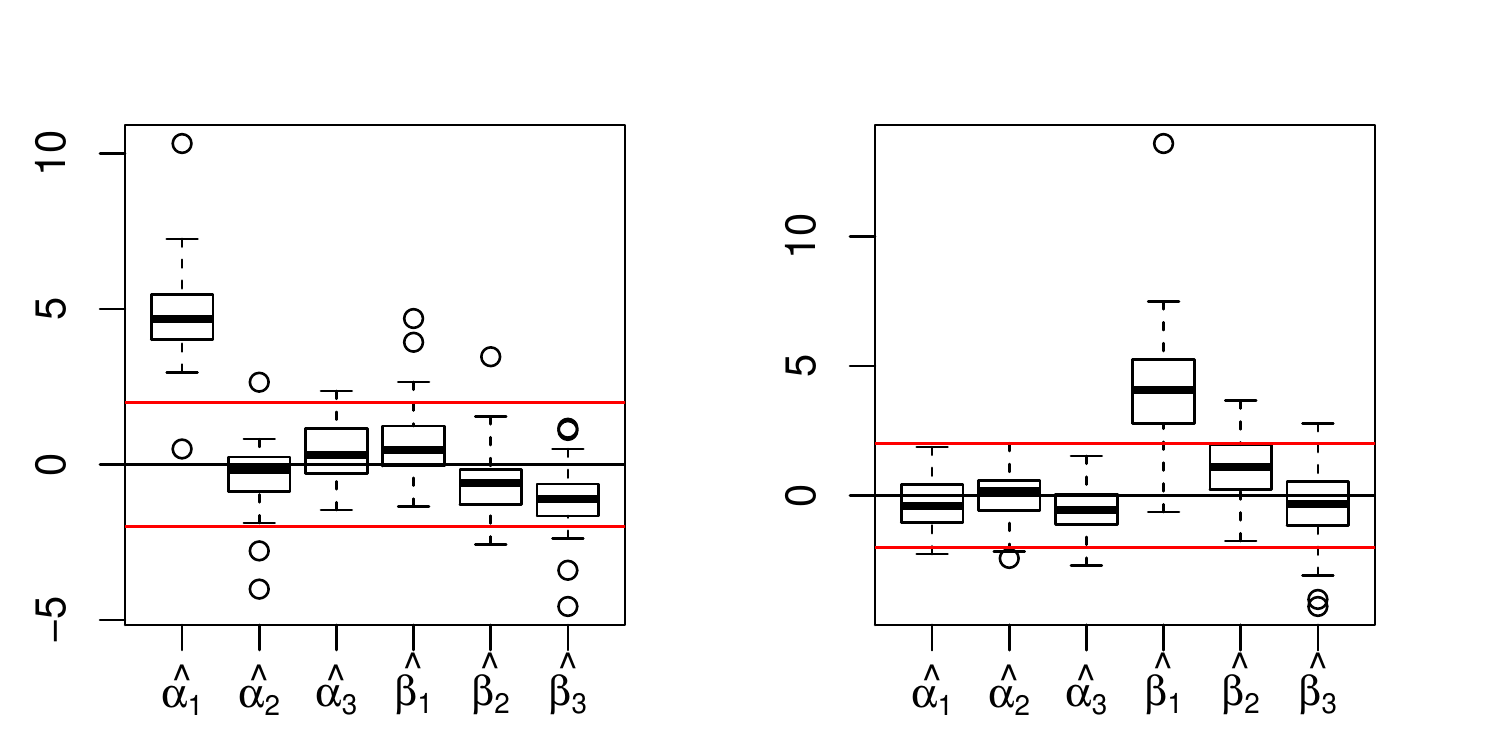}
\end{center}
\caption{(Left)
Boxplots across states of $t$-statistics for
the parameters in the model for $A_t$ as a function of the past. The
horizontal red lines are at $\pm 2$.
Only $\hat \alpha_1$ is
consistently significantly different from zero across states,
suggesting that the times series of at home mobility $A_t$ is a
memory one process.
(Right) Same for $Y_t$. Only $\hat \beta_1$ is
consistently significantly different from zero across states,
suggesting that the deaths times series $Y_t$ is a memory one
process.}
\label{fig::Markov}
\end{figure}

\noindent
{\it 2. The Markov Assumption.}
In Section \ref{subsection::weights},
to estimate the weights, we have made the Markov assumption that
$A_{t-\delta}$ is conditionally
independent of the past given
$(A_{t-1-\delta},L_{t-1-\delta})$.
We also assumed that
$L_t$ is conditionally independent of the past given
$(A_{t-1-\delta},L_{t-1})$. 
To assess this assumption,
we fit the models
\begin{align*}
A_{t-\delta} &= \alpha_0 + \alpha_1 A_{t-1-\delta} + \alpha_2 A_{t-2-\delta} + \alpha_3 A_{t-3-\delta} + 
\beta_1 L_{t-1-\delta} + \beta_2 L_{t-2-\delta} + \beta_3 L_{t-3-\delta} + \epsilon_t\\
L_t &= \alpha_0 + \alpha_1 A_{t-\delta} + \alpha_2 A_{t-\delta-1} + \alpha_3 A_{t-\delta-2} + 
\beta_1 L_{t-1} + \beta_2 L_{t-2} + \beta_3 L_{t-3} + \delta_t.
\end{align*}
Figure \ref{fig::Markov}
shows boxplots of the t-statistics
for these parameters.
The evidence suggests that the first order
Markov assumption is reasonable.
The weak dependence of $A_t$ on past values of $Y_t$ is consistent with the weights $W_t$ 
having almost no effect, i.e. there is
little confounding due to past deaths.
However,
this assessment still assumes
that the Markov assumption is homogeneous, that is,
that the law of
$A_t$ given $(A_{t-1},Y_{t-1})$
is constant over time.
This assumption is not checkable
without invoking further assumptions.

\bigskip

\begin{figure}
\begin{center}
\begin{tabular}{c}
\includegraphics[scale=0.5]{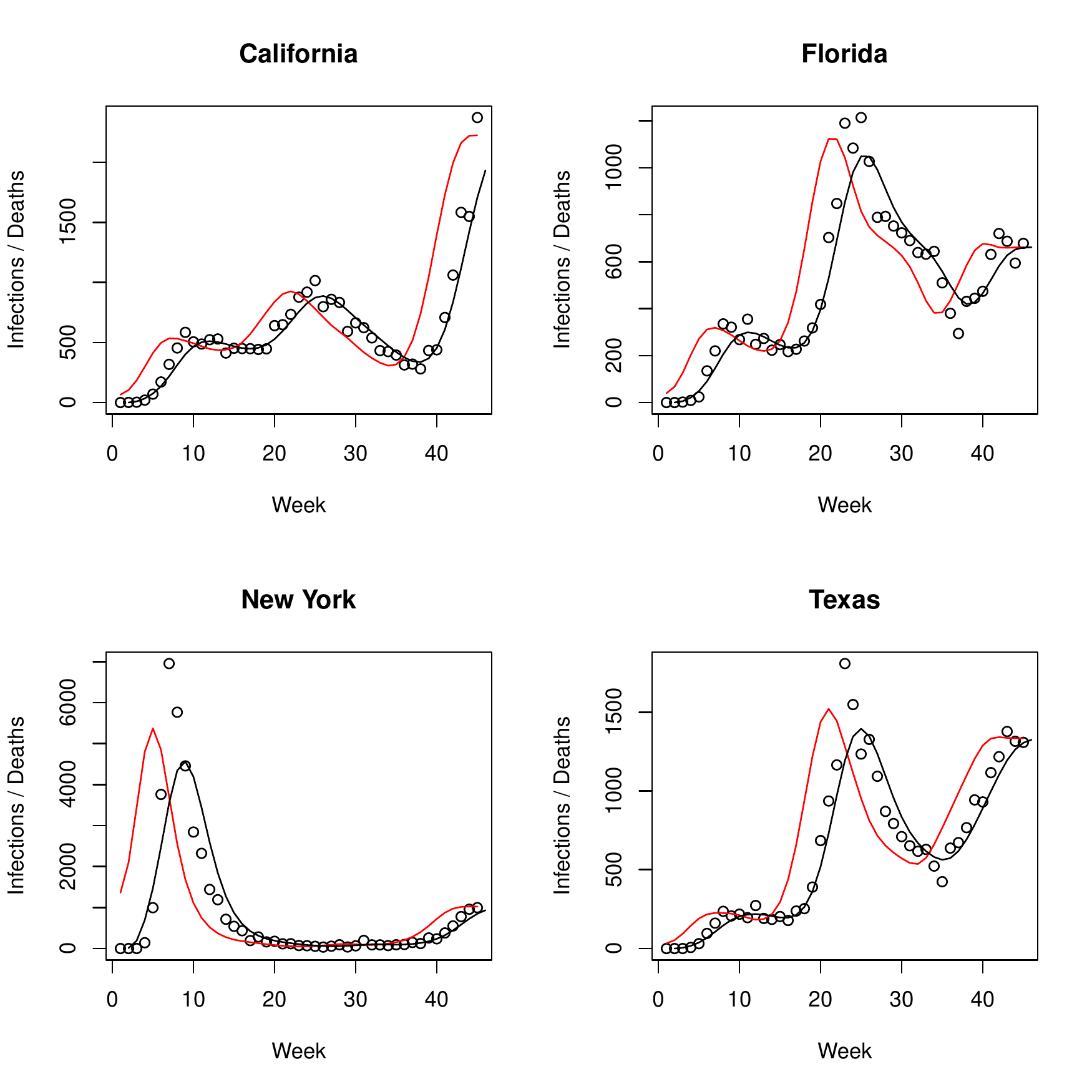}
\end{tabular}
\end{center}
\vspace{-0.2in}
\caption{Inferred infections in four states. The red curve is $\widehat{\tilde I_t}$,
the estimate of the number of infections times the probability of dying if infected by Covid-19,
$\tilde I_t = d(t) I_t$.
The black curve
is deaths $F \widehat{I}$ computed from the optimization with $\lambda  = 1$ in (\ref{eq::lambda}),
and the dots are the observed deaths.}
\label{fig::infections}

\begin{center}
\begin{tabular}{c}
\includegraphics[scale=0.5]{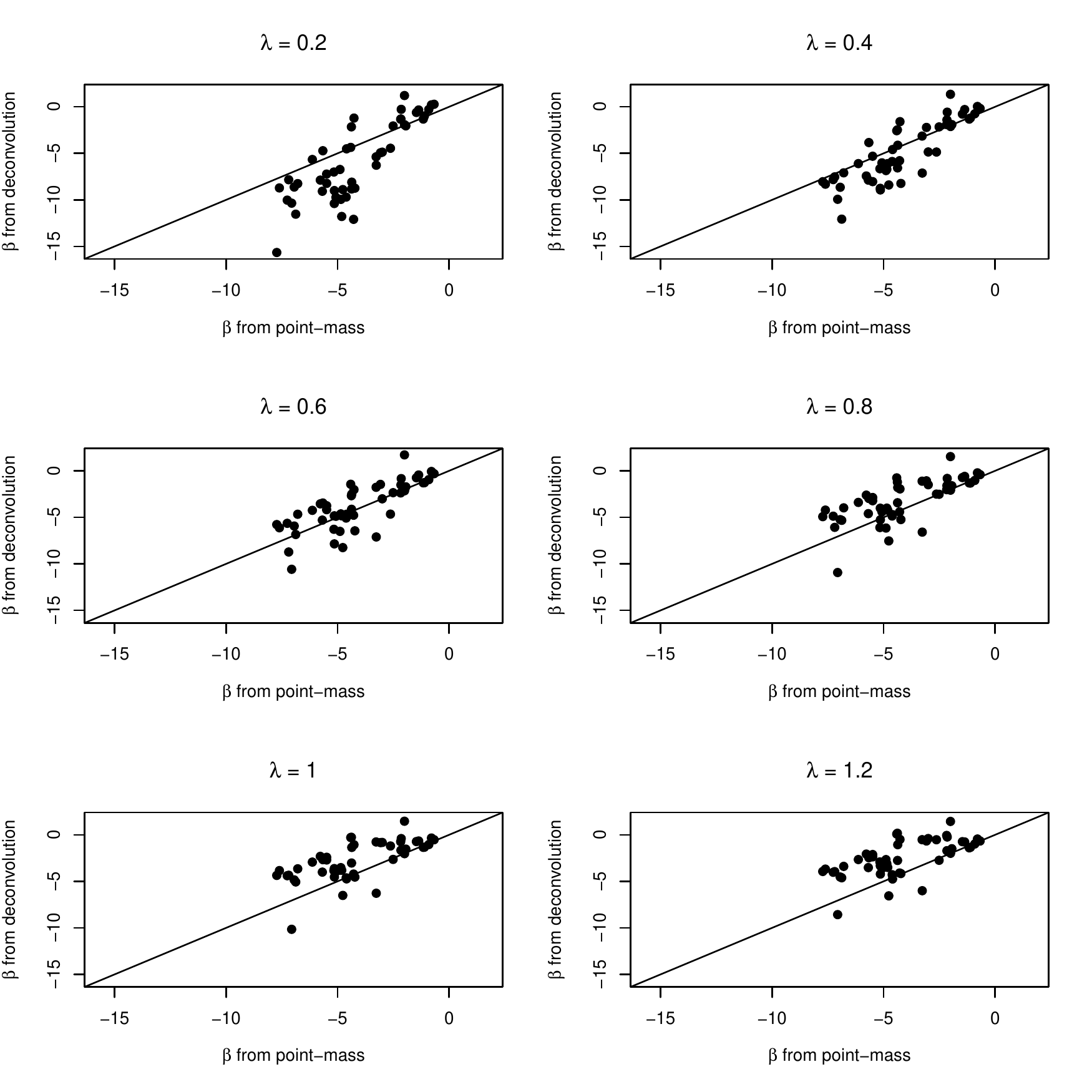}
\end{tabular}
\end{center}
\vspace{-0.2in}
\caption{Comparison of estimates of $\hat\beta$ from the MSM using the 
point-mass approximation versus using estimates of infections via deconvolution for different values of $\lambda$.}
\label{fig::betas_deaths_infections}
\end{figure}

\noindent
{\it 3. Point Mass Versus Deconvolution.}  
Recall that in Section \ref{section::simplified}
we approximated $f_0(s,t)$ with a point mass at $t-\delta$ with $\delta=4$.
An alternative is to
solve the estimating equation using $g$ defined as in
(\ref{eq::general})
but this is numerically very unstable.
Yet another alternative to the point
mass approximation is to estimate the number of infections $I$ by
deconvolution. From the number of infections, we can estimate the
model parameters as in Section \ref{section::fitting} without making the point mass approximation,
using $\log(I)$
as the outcome variable. 
We infer $\tilde{I}_t = d(t) I_t$
from
the optimization:
\begin{equation}
\min_{I \geq 0}\| Y - F \tilde{I} \|_2^2 + \lambda \sum_{r = 2}^{T-1}(\tilde I_r - \tilde I_{r-1})^2,
\label{eq::lambda}
\end{equation}
where $Y$ denotes the vector of weekly deaths and $F$ is a matrix with
$(i,j)$-entry equal to $f(i, j)$ if $j \leq i$ and zero otherwise;
that is, $F_{ij}$ is proportional to the probability of dying at time $j$ given that
infection occurred at time $i$. 
The parameter $\lambda$ is
user-specified and represents a penalty imposed on non-smooth
solutions. Because $f$ is proportional to the density of a Gamma
random variable, we have $F_{ii} = f(i, i) = 0$. To ensure nonzero
elements on the diagonal of $F$, we remove the first row and last
column (all zeros) from $F$ and solve (\ref{eq::lambda}) using $Y = (Y_2,
\ldots, Y_T)$, thus obtaining an estimate of $\tilde I = (\tilde I_1, \ldots,
\tilde I_{T-1})$. To enforce nonnegative values of $I$, we use the
constrained optimization routine \texttt{L-BFGS-B} from \texttt{optim}
in \texttt{R}. Using a penalty $\lambda = 1$, we report the inferred
infections 
(up to proportionality)
$\widehat{I}$ (red line) for California, Florida, New York
and Texas in Figure \ref{fig::infections} along with the implied
deaths computed as $F \widehat{I}$. 
The latter match the observed deaths well, leading credence to this procedure.
In Figure
\ref{fig::betas_deaths_infections}, we compare the estimates of
$\beta$ from the MSM using the point-mass approximation and those from
the MSM using the estimates of infections from the deconvolution
step. 
The estimates
are in rough agreement as
they lie near the diagonal.

\bigskip

\begin{figure}
	\centering
	\begin{subfigure}[t]{.45\textwidth}
		\centering
		\includegraphics[width=\linewidth]{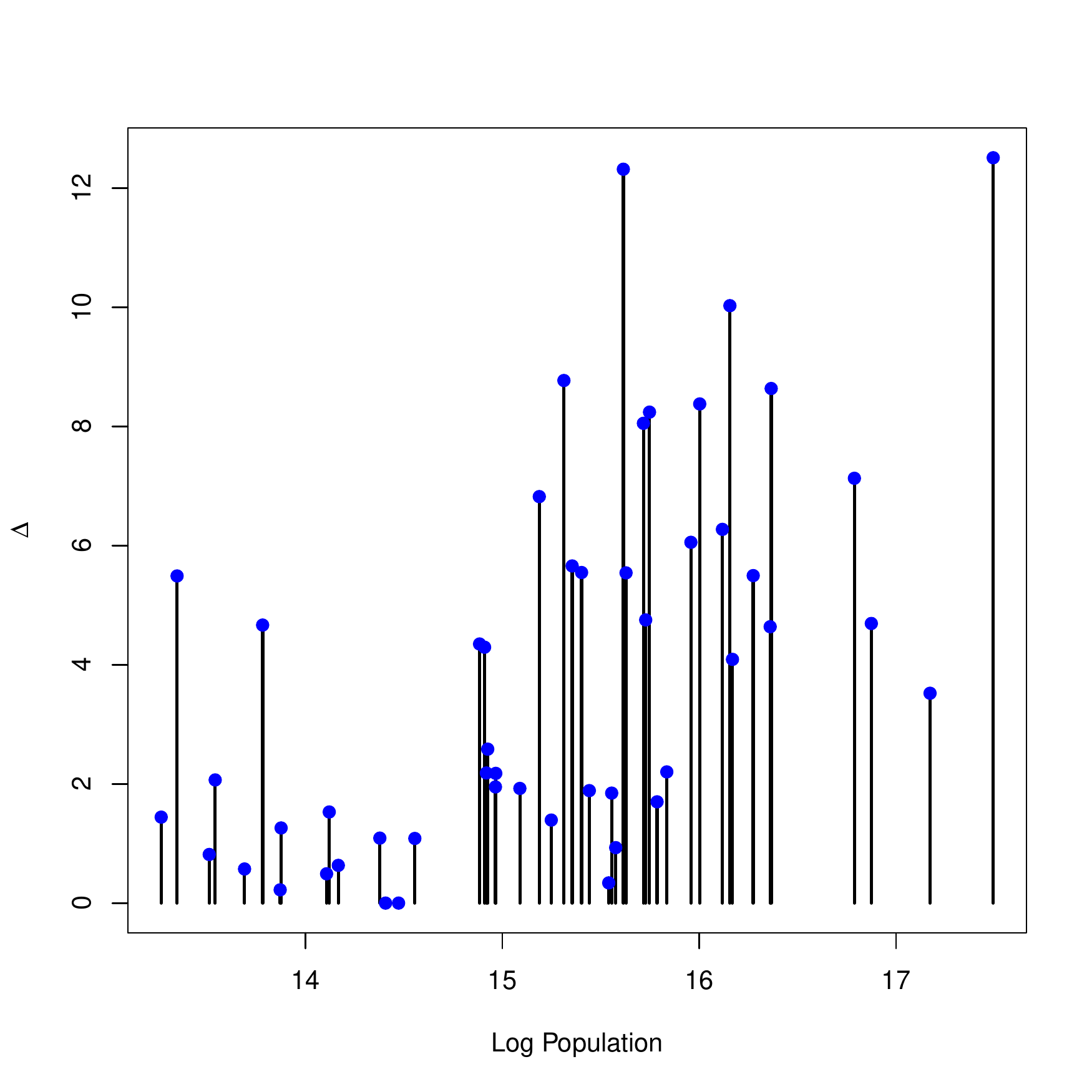}
		\caption{Minimum value of $\Delta$ versus log-population for each state, such that
			unmeasured
			confounding of size $\Delta\, {\rm se}(\hat\beta)$
			causes the
			confidence interval for $\beta$ to contain 0. 
			For most states, it takes
			a fairly large $\Delta$
			to	lose statistical significance.}
 \label{fig::unmeasured}
	\end{subfigure}\hfill
	\begin{subfigure}[t]{.45\textwidth}
		\centering
		\includegraphics[width=\linewidth]{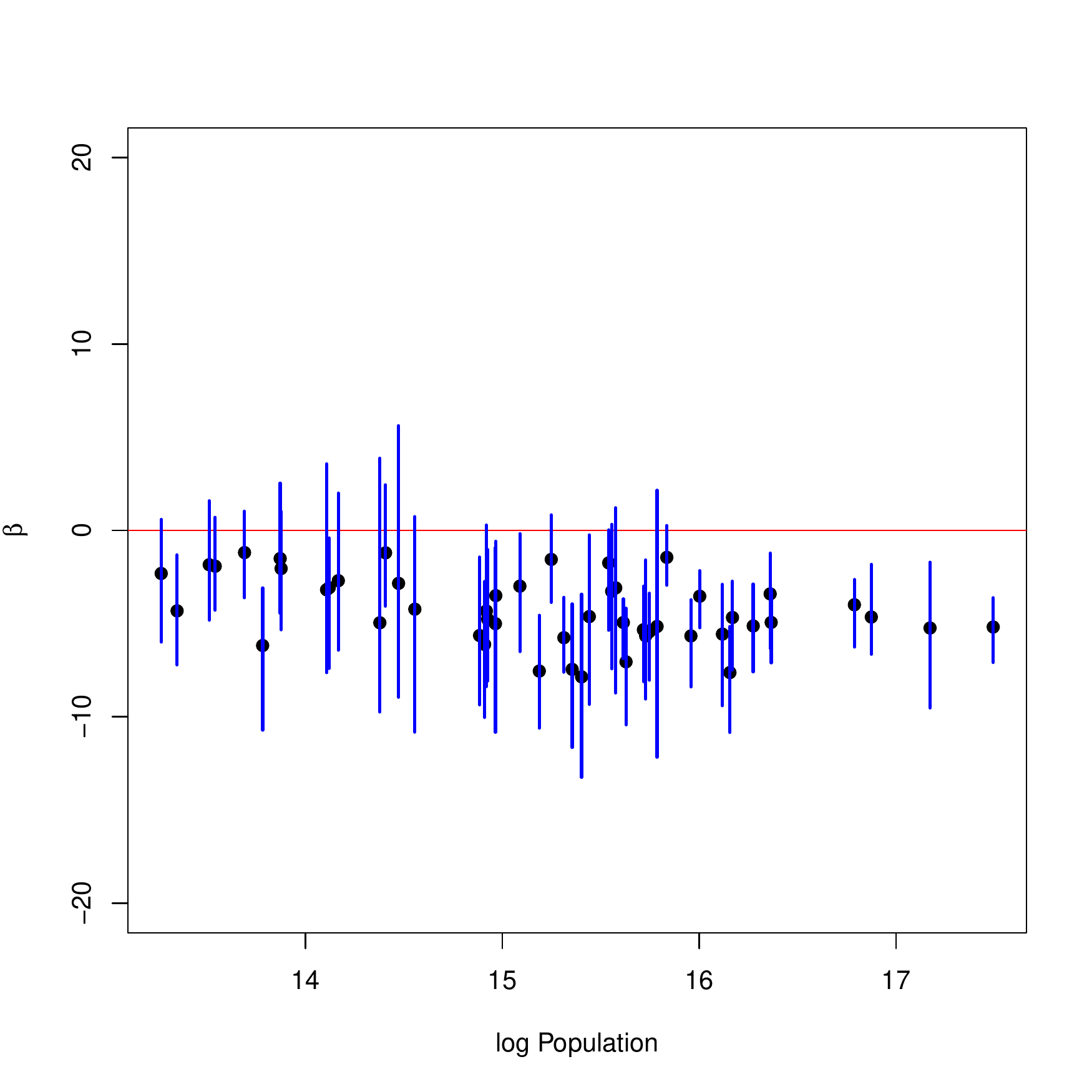}
	\caption{The blue line segments span
		the lower and upper bounds of $\hat\beta$ over the weights $1/\Gamma \leq \tilde W_t/W_t \leq \Gamma$
		with $\Gamma = 3$. 
		The black dots are the original point estimates.
		The effects for most large and medium states remain significant, indicating robustness to unmeasured confounding.	}
	\label{fig::unmeasured2}
	\end{subfigure}
	\caption{Unmeasured confounding sensitivity plots.}
\end{figure}

\noindent
{\it 4. Unmeasured Confounding.}
At time $t$,
we treated
$(A_1,Y_1),\ldots, (A_{t-1},Y_{t-1})$ as
confounders.
Now suppose there is an unmeasured
confounder $U$.
We would like to assess
$|\hat\beta_U - \hat\beta|$
where $\hat\beta_U$ is the value of our estimate
if we had access to $U$.
This quantity is not identified
and so any sensitivity analysis
must invoke some extra assumption.
Let
$\Delta = |\hat\beta_U - \hat\beta|/{\rm se}(\hat\beta)$
denote the unobserved confounding on the standard error scale.
So $\Delta =0$ corresponds
to no unmeasured confounding,
$\Delta =1$ corresponds to saying that
the unmeasured confounding
is the same size as the standard error, etc.
For each state, 
we enlarge the confidence interval by
$\Delta\, {\rm se}(\hat\beta)$.
We can then ask:
how large would $\Delta$ have to be so that
the enlarged confidence interval would contain 0.
Figure \ref{fig::unmeasured}
shows this critical $\Delta$.
We see that for most states, it takes
a fairly large $\Delta$ to 
lose statistical significance.
A substantial number of medium to large states are 
are quite robust to unmeasured confounding.

Adding other potential within state confounders
would be desirable
but, in a within-state analysis,
we can only accommodate time varying
confounders.
(A fixed confounder is a single variable with no replication
and can only be used an across state analysis.)
So far we do not have any within-state
time varying variables that would
be expected to directly affect both $A_t$ and $Y_t$. 
One could imagine that
a variable like ``the percentage of rural cases''
could change over time
and possibly affect both variables
but we do not have such data.

Next we consider a second style of sensitivity analysis
inspired by the approach in 
\cite{rosenbaum2010design}.
The effect of unmeasured confounding
in our analysis is that
the weights $W_t$ are misspecified.
If there are unobserved confounders $U_t$, then the correct
weights are
$$
\tilde{W}_t = \prod_{s=1}^t  \frac{\pi(A_s| \overline{A}_{s-1})}{\pi(A_s| \overline{A}_{s-1},\overline{Y}_{s-1}, \overline{U}_{s-1})}
$$
whereas we estimated the weights
$$
W_t = \prod_{s=1}^t  \frac{\pi(A_s| \overline{A}_{s-1})}{\pi(A_s| \overline{A}_{s-1},\overline{Y}_{s-1})}.
$$
To assess this impact
we find the maximum and minimum $\hat\beta$
under the assumption that
$$
\frac{W_t}{\Gamma} \leq \tilde W_t \leq \Gamma W_t
$$
for $t=1,\ldots, T$
and some $\Gamma \geq 1$.
Similar ideas for static, binary treatments have been
considered in 
\cite{zhao2019sensitivity,
yadlowsky2018bounds}.
Figure \ref{fig::unmeasured2}
shows the bounds on $\hat\beta$ using
$\Gamma = 3$.
Even with this fairly large value of $\Gamma$
the effects for most large and medium states remain significant
indicating robustness to unmeasured confounding.
(The method for computing the bounds is in
\cite{mevw}.)

\subsection{Across Versus Within States}

We have focused on within state estimation.
An alternative is to fit a model
across states as well.
Although we are skeptical of
combining data over states
we do so here for completeness.
We fit the blip model with common $\beta$ and, rather than
include state level covariates such as population size, proportion of residents in cities, etc., we use
a fixed effect for each state.
The resulting estimates of $\beta$ and standard errors
for $k=1,2,3,4$ are:

\begin{center}
\begin{tabular}{lcc}
$k$ & $\hat\beta$ & standard error\\ \hline
1   & -5.20 & 0.27\\
2   & -4.60 & 0.27\\
3   & -3.82 & 0.34\\
4   & -2.83 & 0.43
\end{tabular}
\end{center}

The estimates are consistent with the within state models.
AIC chooses $k=1$, which conflicts with the within state analysis with favors larger $k$
for larger states. 
The likely reason is that combining states adds variability in the combined
dataset since $\beta$'s and $\nu(t)$'s are different between states, so there 
is less signal compared to the noise to estimate a more complicated relationship than a linear.
A natural extension of this model is to use a random effects approach, although we do not pursue that here.

\section{Discussion}
\label{section::conclusion}

Our approach
to modeling the causal effect
of mobility on deaths
is to construct
a marginal structural model
whose parameters are estimated
by solving an
estimating equation.
We model each state separately
to reduce confounding
due to state differences.
Our approach has several 
advantages and disadvantages.

Our modeling assumptions are reasonable
in the short term but
not in the long term.
Eventually, the effects of
acquired immunity, masks, vaccinations etc
might have to be accounted for
by using a more complex form of $\nu$.
\red{Also, the effect of mobility $\beta$ could change
with new variants.}


Estimating the model parameters
comes down to
solving the estimating equation
(\ref{eq::thisisit}).
Computing standard errors
and confidence intervals is then
straightforward.
This is in contrast to more
traditional and Icarian epidemic modeling
which requires estimating
many parameters
using grid searches or MCMC.
Provably valid
confidence intervals are elusive
for those methods.
On the other hand,
the more detailed models
might be more realistic
and can capture effects
that our simple model cannot capture.
Moreover, our inferences
are asymptotic in nature.
When comparing exact Bayesian methods
to approximate frequentist methods
it is hard to argue that one approach is
more valid than the other.

We believe that
focusing on weekly data
at the state level
gives us the best
chance of getting data
of reasonable quality
and helps avoid confounding related to
state differences.
Further, this allows the causal effect
to vary between states.
But this results
in a paucity of data,
a few dozen observations per state.
This limits the complexity
of the models we can fit
and it requires that we make
a homogeneous Markov assumption.
A natural compromise 
worthy of future investigation would be
to use some sort of random effects
model to allow modeling all states
simultaneously.
This could also permit
using data from other countries.
At any rate, there is a tradeoff:
within state analysis requires stronger modeling
assumptions while
analyzing all states together requires
assuming independence and it assumes we can
model all sources of between state confounding.

Detailed dynamic modeling
versus the
more traditional
causal modeling
done here (and in
\cite{chernozhukov2020causal})
represent two different
approaches to
causal inference for epidemics.
It would be interesting to see 
a general comparison of these approaches, 
perhaps eventually leading to some sort of fusion
of these ideas.

\red{
Finally, let us recap
the null paradox.
Any nonlinear, sequentially specified parametric
model ---
which includes most epidemic models ---
has the following problem.
There is no value of the parameters
that allows both (i) the outcome is conditionally
dependent on the intervention variable $A$ and
(ii) there is no causal effect of $A$.
But, due to baseline variables $U$,
(i) and (ii) can both be true.
This means that we would find a causal effect even if there is no such effect.
We can in principle avoid 
the null paradox by
using nonparametric models
but then the model complexity explodes as $T$ increases
leading to the curse of dimensionality.
Linear models
avoid the null paradox but
caution is still needed since
the causal effect $\psi(a)$
involves complicated nonlinear functions of
the regression parameters.
Hence, the model is very difficult to interpret
and the individual regression parameters do not
have a causal interpretation.
Also, most epidemic models are not linear.
}

The quickly growing literature on
using sequentially specified epidemic models
does include such models.
MSMs avoid the null paradox, 
and this is another reason for using MSMs
(or some other semiparametric causal model
such as structural nested models).
In our case
we motivated the MSM by
starting with a 
sequentially specified model.
This seems like a reasonable
approach for using epidemic models
to define an MSM but there may be other
approaches as well.

\section*{Acknowledgments}
The authors would like to thank Rob Tibshirani
and the reviewers
for providing helpful feedback
on an earlier draft of the paper.  
Edward Kennedy gratefully acknowledges support from NSF Grant DMS1810979.

\begin{supplement}
\stitle{Supplement A: Plots for all states.}
\slink[doi]{}
\sdescription{Plots of the data and counterfactual curves for all states.}
\end{supplement}

\begin{supplement}
\stitle{Supplement B: AIC plots.}
\slink[doi]{}
\sdescription{Plots of the value of $k$ selected by AIC.}
\end{supplement}

\begin{supplement}
\stitle{Supplement C: Deconvolution.}
\slink[doi]{}
\sdescription{Plots of the deconvolved data for all states.}
\end{supplement}

\bibliographystyle{imsart-nameyear} 
\bibliography{paper.bib}

\begin{thebibliography}{26}

\bibitem[\protect\citeauthoryear{Bj{\o}rnstad}{2018}]{bjornstad2018epidemics}
\begin{barticle}[author]
\bauthor{\bsnm{Bj{\o}rnstad},~\bfnm{Ottar~N}\binits{O.~N.}}
(\byear{2018}).
\btitle{Epidemics}.
\bjournal{Models and data using R: Springer International Publishing}
\bpages{318}.
\end{barticle}
\endbibitem

\bibitem[\protect\citeauthoryear{Bonvini et~al.}{2021}]{mevw}
\begin{barticle}[author]
\bauthor{\bsnm{Bonvini},~\bfnm{M.}\binits{M.}},
  \bauthor{\bsnm{Kennedy},~\bfnm{E.}\binits{E.}},
  \bauthor{\bsnm{Ventura},~\bfnm{V.}\binits{V.}} \AND
  \bauthor{\bsnm{Wasserman},~\bfnm{L.}\binits{L.}}
(\byear{2021}).
\btitle{Propensity Scores and Sensivity Analysis for Marginal Structural Models
  with Continuous Treatments}.
\bjournal{In preparation}.
\end{barticle}
\endbibitem

\bibitem[\protect\citeauthoryear{Brauer, Castillo-Chavez and
  Castillo-Chavez}{2012}]{brauer2012mathematical}
\begin{bbook}[author]
\bauthor{\bsnm{Brauer},~\bfnm{Fred}\binits{F.}},
  \bauthor{\bsnm{Castillo-Chavez},~\bfnm{Carlos}\binits{C.}} \AND
  \bauthor{\bsnm{Castillo-Chavez},~\bfnm{Carlos}\binits{C.}}
(\byear{2012}).
\btitle{Mathematical models in population biology and epidemiology}
\bvolume{2}.
\bpublisher{Springer}.
\end{bbook}
\endbibitem

\bibitem[\protect\citeauthoryear{Chang et~al.}{2020}]{chang2020mobility}
\begin{barticle}[author]
\bauthor{\bsnm{Chang},~\bfnm{Serina}\binits{S.}},
  \bauthor{\bsnm{Pierson},~\bfnm{Emma}\binits{E.}},
  \bauthor{\bsnm{Koh},~\bfnm{Pang~Wei}\binits{P.~W.}},
  \bauthor{\bsnm{Gerardin},~\bfnm{Jaline}\binits{J.}},
  \bauthor{\bsnm{Redbird},~\bfnm{Beth}\binits{B.}},
  \bauthor{\bsnm{Grusky},~\bfnm{David}\binits{D.}} \AND
  \bauthor{\bsnm{Leskovec},~\bfnm{Jure}\binits{J.}}
(\byear{2020}).
\btitle{Mobility network models of COVID-19 explain inequities and inform
  reopening}.
\bjournal{Nature}
\bpages{1--6}.
\end{barticle}
\endbibitem

\bibitem[\protect\citeauthoryear{Chernozhukov, Kasahara and
  Schrimpf}{2020}]{chernozhukov2020causal}
\begin{barticle}[author]
\bauthor{\bsnm{Chernozhukov},~\bfnm{Victor}\binits{V.}},
  \bauthor{\bsnm{Kasahara},~\bfnm{Hiroyuki}\binits{H.}} \AND
  \bauthor{\bsnm{Schrimpf},~\bfnm{Paul}\binits{P.}}
(\byear{2020}).
\btitle{Causal impact of masks, policies, behavior on early COVID-19 pandemic
  in the US}.
\bjournal{arXiv preprint arXiv:2005.14168}.
\end{barticle}
\endbibitem

\bibitem[\protect\citeauthoryear{Fong et~al.}{2018}]{fong2018covariate}
\begin{barticle}[author]
\bauthor{\bsnm{Fong},~\bfnm{Christian}\binits{C.}},
  \bauthor{\bsnm{Hazlett},~\bfnm{Chad}\binits{C.}},
  \bauthor{\bsnm{Imai},~\bfnm{Kosuke}\binits{K.}} \betal{et~al.}
(\byear{2018}).
\btitle{Covariate balancing propensity score for a continuous treatment:
  Application to the efficacy of political advertisements}.
\bjournal{The Annals of Applied Statistics}
\bvolume{12}
\bpages{156--177}.
\end{barticle}
\endbibitem

\bibitem[\protect\citeauthoryear{IHME}{2020}]{ihme2020modeling}
\begin{barticle}[author]
\bauthor{\bsnm{IHME}}
(\byear{2020}).
\btitle{Modeling COVID-19 scenarios for the United States}.
\bjournal{Nature Medicine}.
\end{barticle}
\endbibitem

\bibitem[\protect\citeauthoryear{Kermack and
  McKendrick}{1927}]{kermack1927contribution}
\begin{barticle}[author]
\bauthor{\bsnm{Kermack},~\bfnm{William~Ogilvy}\binits{W.~O.}} \AND
  \bauthor{\bsnm{McKendrick},~\bfnm{Anderson~G}\binits{A.~G.}}
(\byear{1927}).
\btitle{A contribution to the mathematical theory of epidemics}.
\bjournal{Proceedings of the royal society of london. Series A, Containing
  papers of a mathematical and physical character}
\bvolume{115}
\bpages{700--721}.
\end{barticle}
\endbibitem

\bibitem[\protect\citeauthoryear{Liao and Meyer}{2018}]{cgam}
\begin{barticle}[author]
\bauthor{\bsnm{Liao},~\bfnm{X.}\binits{X.}} \AND
  \bauthor{\bsnm{Meyer},~\bfnm{M.}\binits{M.}}
(\byear{2018}).
\btitle{cgam: Constrained generalized additive model}.
\bjournal{xxxx}
\bvolume{xx}
\bpages{xxxx}.
\end{barticle}
\endbibitem

\bibitem[\protect\citeauthoryear{Meyer et~al.}{2008}]{meyer2008inference}
\begin{barticle}[author]
\bauthor{\bsnm{Meyer},~\bfnm{Mary~C}\binits{M.~C.}} \betal{et~al.}
(\byear{2008}).
\btitle{Inference using shape-restricted regression splines}.
\bjournal{The Annals of Applied Statistics}
\bvolume{2}
\bpages{1013--1033}.
\end{barticle}
\endbibitem

\bibitem[\protect\citeauthoryear{Meyer et~al.}{2018}]{meyer2018framework}
\begin{barticle}[author]
\bauthor{\bsnm{Meyer},~\bfnm{Mary~C}\binits{M.~C.}} \betal{et~al.}
(\byear{2018}).
\btitle{A framework for estimation and inference in generalized additive models
  with shape and order restrictions}.
\bjournal{Statistical Science}
\bvolume{33}
\bpages{595--614}.
\end{barticle}
\endbibitem

\bibitem[\protect\citeauthoryear{Neugebauer and van~der
  Laan}{2007}]{neugebauer2007nonparametric}
\begin{barticle}[author]
\bauthor{\bsnm{Neugebauer},~\bfnm{Romain}\binits{R.}} \AND
  \bauthor{\bparticle{van~der} \bsnm{Laan},~\bfnm{Mark}\binits{M.}}
(\byear{2007}).
\btitle{Nonparametric causal effects based on marginal structural models}.
\bjournal{Journal of Statistical Planning and Inference}
\bvolume{137}
\bpages{419--434}.
\end{barticle}
\endbibitem

\bibitem[\protect\citeauthoryear{Robins}{1986}]{robins1986new}
\begin{barticle}[author]
\bauthor{\bsnm{Robins},~\bfnm{James}\binits{J.}}
(\byear{1986}).
\btitle{A new approach to causal inference in mortality studies with a
  sustained exposure period—application to control of the healthy worker
  survivor effect}.
\bjournal{Mathematical modelling}
\bvolume{7}
\bpages{1393--1512}.
\end{barticle}
\endbibitem

\bibitem[\protect\citeauthoryear{Robins}{1989}]{robins1989analysis}
\begin{barticle}[author]
\bauthor{\bsnm{Robins},~\bfnm{James~M}\binits{J.~M.}}
(\byear{1989}).
\btitle{The analysis of randomized and non-randomized AIDS treatment trials
  using a new approach to causal inference in longitudinal studies}.
\bjournal{Health service research methodology: a focus on AIDS}
\bpages{113--159}.
\end{barticle}
\endbibitem

\bibitem[\protect\citeauthoryear{Robins}{2000}]{robins2000marginal2}
\begin{bincollection}[author]
\bauthor{\bsnm{Robins},~\bfnm{James~M}\binits{J.~M.}}
(\byear{2000}).
\btitle{Marginal structural models versus structural nested models as tools for
  causal inference}.
In \bbooktitle{Statistical models in epidemiology, the environment, and
  clinical trials}
\bpages{95--133}.
\bpublisher{Springer}.
\end{bincollection}
\endbibitem

\bibitem[\protect\citeauthoryear{Robins, Hernan and
  Brumback}{2000}]{robins2000marginal}
\begin{bmisc}[author]
\bauthor{\bsnm{Robins},~\bfnm{James~M}\binits{J.~M.}},
  \bauthor{\bsnm{Hernan},~\bfnm{Miguel~Angel}\binits{M.~A.}} \AND
  \bauthor{\bsnm{Brumback},~\bfnm{Babette}\binits{B.}}
(\byear{2000}).
\btitle{Marginal structural models and causal inference in epidemiology}.
\end{bmisc}
\endbibitem

\bibitem[\protect\citeauthoryear{Robins and Wasserman}{1997}]{rw}
\begin{bincollection}[author]
\bauthor{\bsnm{Robins},~\bfnm{James~M}\binits{J.~M.}} \AND
  \bauthor{\bsnm{Wasserman},~\bfnm{L}\binits{L.}}
(\byear{1997}).
\btitle{Estimation of Effects of Sequential Treatments by Reparameterizing
  Directed Acyclic Graphs}.
In \bbooktitle{Proceedings of the Thirteenth Conference on Uncertainty in
  Artificial Intelligence}
\bpages{409-420}.
\bpublisher{Morgan Kaufmann}.
\end{bincollection}
\endbibitem

\bibitem[\protect\citeauthoryear{Rosenbaum et~al.}{2010}]{rosenbaum2010design}
\begin{bbook}[author]
\bauthor{\bsnm{Rosenbaum},~\bfnm{Paul~R}\binits{P.~R.}} \betal{et~al.}
(\byear{2010}).
\btitle{Design of observational studies}
\bvolume{10}.
\bpublisher{Springer}.
\end{bbook}
\endbibitem

\bibitem[\protect\citeauthoryear{Shi and Ban}{2020}]{shi2020capping}
\begin{barticle}[author]
\bauthor{\bsnm{Shi},~\bfnm{Yunfeng}\binits{Y.}} \AND
  \bauthor{\bsnm{Ban},~\bfnm{Xuegang}\binits{X.}}
(\byear{2020}).
\btitle{Capping Mobility to Control COVID-19: A Collision-based Infectious
  Disease Transmission Model}.
\bjournal{medRxiv}.
\end{barticle}
\endbibitem

\bibitem[\protect\citeauthoryear{Tsiatis}{2007}]{tsiatis2007semiparametric}
\begin{bbook}[author]
\bauthor{\bsnm{Tsiatis},~\bfnm{Anastasios}\binits{A.}}
(\byear{2007}).
\btitle{Semiparametric theory and missing data}.
\bpublisher{Springer Science \& Business Media}.
\end{bbook}
\endbibitem

\bibitem[\protect\citeauthoryear{Unwin et~al.}{2020}]{unwin2020state}
\begin{barticle}[author]
\bauthor{\bsnm{Unwin},~\bfnm{H~Juliette~T}\binits{H.~J.~T.}},
  \bauthor{\bsnm{Mishra},~\bfnm{Swapnil}\binits{S.}},
  \bauthor{\bsnm{Bradley},~\bfnm{Valerie~C}\binits{V.~C.}},
  \bauthor{\bsnm{Gandy},~\bfnm{Axel}\binits{A.}},
  \bauthor{\bsnm{Mellan},~\bfnm{Thomas~A}\binits{T.~A.}},
  \bauthor{\bsnm{Coupland},~\bfnm{Helen}\binits{H.}},
  \bauthor{\bsnm{Ish-Horowicz},~\bfnm{Jonathan}\binits{J.}},
  \bauthor{\bsnm{Vollmer},~\bfnm{Michaela~AC}\binits{M.~A.}},
  \bauthor{\bsnm{Whittaker},~\bfnm{Charles}\binits{C.}},
  \bauthor{\bsnm{Filippi},~\bfnm{Sarah~L}\binits{S.~L.}} \betal{et~al.}
(\byear{2020}).
\btitle{State-level tracking of COVID-19 in the United States}.
\bjournal{Nature communications}
\bvolume{11}
\bpages{1--9}.
\end{barticle}
\endbibitem

\bibitem[\protect\citeauthoryear{Vansteelandt
  et~al.}{2014}]{vansteelandt2014structural}
\begin{barticle}[author]
\bauthor{\bsnm{Vansteelandt},~\bfnm{Stijn}\binits{S.}},
  \bauthor{\bsnm{Joffe},~\bfnm{Marshall}\binits{M.}} \betal{et~al.}
(\byear{2014}).
\btitle{Structural nested models and G-estimation: the partially realized
  promise}.
\bjournal{Statistical Science}
\bvolume{29}
\bpages{707--731}.
\end{barticle}
\endbibitem

\bibitem[\protect\citeauthoryear{Xiong et~al.}{2020}]{xiong2020mobile}
\begin{barticle}[author]
\bauthor{\bsnm{Xiong},~\bfnm{Chenfeng}\binits{C.}},
  \bauthor{\bsnm{Hu},~\bfnm{Songhua}\binits{S.}},
  \bauthor{\bsnm{Yang},~\bfnm{Mofeng}\binits{M.}},
  \bauthor{\bsnm{Luo},~\bfnm{Weiyu}\binits{W.}} \AND
  \bauthor{\bsnm{Zhang},~\bfnm{Lei}\binits{L.}}
(\byear{2020}).
\btitle{Mobile device data reveal the dynamics in a positive relationship
  between human mobility and COVID-19 infections}.
\bjournal{Proceedings of the National Academy of Sciences}
\bvolume{117}
\bpages{27087--27089}.
\end{barticle}
\endbibitem

\bibitem[\protect\citeauthoryear{Yadlowsky et~al.}{2018}]{yadlowsky2018bounds}
\begin{barticle}[author]
\bauthor{\bsnm{Yadlowsky},~\bfnm{Steve}\binits{S.}},
  \bauthor{\bsnm{Namkoong},~\bfnm{Hongseok}\binits{H.}},
  \bauthor{\bsnm{Basu},~\bfnm{Sanjay}\binits{S.}},
  \bauthor{\bsnm{Duchi},~\bfnm{John}\binits{J.}} \AND
  \bauthor{\bsnm{Tian},~\bfnm{Lu}\binits{L.}}
(\byear{2018}).
\btitle{Bounds on the conditional and average treatment effect with unobserved
  confounding factors}.
\bjournal{arXiv preprint arXiv:1808.09521}.
\end{barticle}
\endbibitem

\bibitem[\protect\citeauthoryear{Zhao, Small and
  Bhattacharya}{2019}]{zhao2019sensitivity}
\begin{barticle}[author]
\bauthor{\bsnm{Zhao},~\bfnm{Qingyuan}\binits{Q.}},
  \bauthor{\bsnm{Small},~\bfnm{Dylan~S}\binits{D.~S.}} \AND
  \bauthor{\bsnm{Bhattacharya},~\bfnm{Bhaswar~B}\binits{B.~B.}}
(\byear{2019}).
\btitle{Sensitivity analysis for inverse probability weighting estimators via
  the percentile bootstrap}.
\bjournal{Journal of the Royal Statistical Society: Series B (Statistical
  Methodology)}
\bvolume{81}
\bpages{735--761}.
\end{barticle}
\endbibitem

\bibitem[\protect\citeauthoryear{Zhou and Wodtke}{2018}]{zhou2018residual}
\begin{barticle}[author]
\bauthor{\bsnm{Zhou},~\bfnm{Xiang}\binits{X.}} \AND
  \bauthor{\bsnm{Wodtke},~\bfnm{Geoffrey~T}\binits{G.~T.}}
(\byear{2018}).
\btitle{Residual balancing weights for marginal structural models: with
  application to analyses of time-varying treatments and causal mediation}.
\bjournal{arXiv preprint arXiv:1807.10869}.
\end{barticle}
\endbibitem

\end{thebibliography}

\end{document}